\documentclass[11pt]{article}
\usepackage{array}
\usepackage{dsfont}
\usepackage{wrapfig}
\usepackage[small, it]{caption}
\usepackage{latexsym} 
\usepackage{amsmath, mathtools}
\usepackage{amssymb, amsthm}
\usepackage{graphicx}
\usepackage[english]{babel}
\usepackage{layout}
\usepackage{fancyhdr}
\usepackage{slashed}	
\usepackage{bbm}
\usepackage{tabularx}
\usepackage[usenames]{color}
\usepackage[all]{xy}

\makeatletter
\newif\if@preliminary
\@preliminaryfalse
\def\preliminary{\@preliminaryfalse}

\def\bq{\begin{equation}}
\def\eq{\end{equation}}
\def\ba{\begin{eqnarray}}
\def\ea{\end{eqnarray}}

\def\preprintno#1{\def\@preprintno{#1}}
\def\address#1{\def\@address{#1}}
\def\email#1#2{\thanks{\tt #1@{}#2}}
\def\abstract#1{\def\@abstract{#1}}
\renewcommand\abstractname{ABSTRACT}
\newlength\preprintnoskip
\newlength\abstractwidth
\@titlepagetrue
\renewcommand\maketitle{\begin{titlepage}
  \let\footnotesize\small
  \hfill\parbox{\preprintnoskip}{
  \begin{flushright}\@preprintno\end{flushright}}\hspace*{1cm}
  \vskip 60\p@
  \begin{center}
    {\Large\bf\boldmath \@title \par}\vskip 1cm
    {\sc\@author \par}\vskip 3mm
    {\@address \par}
    \if@preliminary
      \vskip 2cm {\large\sf PRELIMINARY DRAFT \par \@date}
    \fi
  \end{center}\par
  \@thanks
  \vfill
  \begin{center}
    \parbox{\abstractwidth}{\centerline{\abstractname}
    \vskip 3mm
    \@abstract}
  \end{center}
  \end{titlepage}
  \setcounter{footnote}{0}
  \let\thanks\relax\let\maketitle\relax
  \gdef\@thanks{}\gdef\@author{}\gdef\@address{}
  \gdef\@title{}\gdef\@abstract{}\gdef\@preprintno{}
}

\def\@citex[#1]#2{\if@filesw\immediate\write\@auxout{\string\citation{#2}}\fi
  \def\@citea{}\@cite{\@for\@citeb:=#2\do
    {\@citea\def\@citea{,\penalty\@m}\@ifundefined
       {b@\@citeb}{{\bf ?}\@warning
       {Citation `\@citeb' on page \thepage \space undefined}}%
\hbox{\csname b@\@citeb\endcsname}}}{#1}}
\def\citerange{\@ifnextchar [{\@tempswatrue\@citexr}{\@tempswafalse\@citexr[]}}
\def\@citexr[#1]#2{\if@filesw\immediate\write\@auxout{\string\citation{#2}}\fi
  \def\@citea{}\@cite{\@for\@citeb:=#2\do
    {\@citea\def\@citea{--\penalty\@m}\@ifundefined
       {b@\@citeb}{{\bf ?}\@warning
       {Citation `\@citeb' on page \thepage \space undefined}}%
\hbox{\csname b@\@citeb\endcsname}}}{#1}}
\long\def\@makecaption#1#2{
  \vskip\abovecaptionskip
  \sbox\@tempboxa{#1: \emph{#2}}
  \ifdim \wd\@tempboxa >\hsize
    #1: \emph{#2}\par
  \else
    \hbox to\hsize{\hfil\box\@tempboxa\hfil}
  \fi
  \vskip\belowcaptionskip}

\newcommand{\GeV}{{\ensuremath\rm GeV}}
\newcommand{\TeV}{{\ensuremath\rm TeV}}

\begin{document}



\preprintno{DESY-12-083  \\[0.5\baselineskip] June 11, 2012} 

\title{Extracting Gluino endpoints\\with event topology patterns}

\author{
  N.~Pietsch\email{niklas.pietsch}{desy.de}$^{\,a}$, 
  J.~Reuter\email{juergen.reuter}{desy.de}$^{\,b}$, 
  K.~Sakurai\email{kazuki.sakurai}{desy.de}$^{\,b}$ 
 and D.~Wiesler\email{daniel.wiesler}{desy.de}$^{\,b}$
}

\address{\it$^a$University of Hamburg, Luruper Chaussee 149, D--22761 Hamburg, Germany\\$^b$DESY Theory Group, Notkestr. 85, D--22603 Hamburg, Germany}

\vspace{-5cm}
\abstract{
In this paper we study the gluino dijet mass edge measurement at the
LHC in a realistic situation including both SUSY and
combinatorical backgrounds together with effects of initial and final state
radiation as well as a finite detector resolution. Three benchmark
scenarios are examined in which the dominant SUSY production process
and also the decay modes are different. 
Several new kinematical variables are proposed to minimize the impact
of SUSY and combinatorial backgrounds 
in the measurement. By selecting events with a particular number of
jets and leptons, we attempt to measure two distinct gluino dijet mass
edges originating from wino $\tilde g \to jj \tilde W$ and bino $\tilde g \to jj \tilde B$ decay modes, separately. We determine the endpoints of distributions of
proposed and existing variables and show that those two edges can be
disentangled and measured within good accuracy, irrespective of the
presence of ISR, FSR, and detector effects. 
}

\maketitle
   
\section{Introduction}

The Large Hadron Collider (LHC) has entered an exciting era 
by seeing a tantalizing excess of Higgs-like events in the mass region around 125\,GeV.
The Higgs boson mass parameter receives a large quantum correction of
the order of a cut off scale and hence new physics that stabilizes the
weak scale is anticipated to be seen in LHC events.   

Supersymmetry (SUSY) is one of the most promising candidates of such new physics models.  
The Minimal Supersymmetric extension of the Standard Model (MSSM)
allows an exact unification of all three forces in the Standard Model
(SM), indicating a grand unified theory at a very high energy scale.  
The lightest SUSY particle (LSP) is stable because of a discrete
symmetry in the MSSM, called R-parity, and can be a viable dark matter
candidate if it is neutral. The R-parity also makes SUSY events in a
collider distinct from the SM background. It implies SUSY particles to
be produced in pairs, with each decaying into the LSP through a
cascade decay chain, leading to multiple jets, leptons and large
missing energy.  The LHC experimental collaborations so far have put 
great effort into searching for a sign of Supersymmetry at the LHC.    
If Supersymmetry is discovered, the next important task is measuring the properties of SUSY particles.  SUSY events contain two LSPs in the final states, which escape detection.
In hadron colliders, the only information we can deduce on the LSP momenta is a vector sum of their transverse momenta, on the basis of the assumption that there are no extra missing particles, such as neutrinos, in the event.  
This makes any measurement about SUSY particles non-trivial and challenging.

Fortunately, many ideas have already been put forward to address this obstacle 
(see \cite{Barr:2010zj} for a review).
The most traditional method is to look for kinematical edges in various invariant mass distributions of the daughter particles 
\cite{Hinchliffe:1996iu}.  
The locations of these edges reveal information on the unknown intermediate particle masses in the decay chain.
Another approach is to use the family of $M_{T2}$-based kinematic variables 
\cite{Lester:1999tx,Lester:2007fq,Nojiri:2008hy,Tovey:2008ui,Serna:2008zk}, 
which often serves the event-by-event best lower bound on the unknown particle mass of interest.
The third option is the polynomial method 
\cite{Desch:2003vw}, 
which attempts to determine all the missing momenta in the event by
solving the kinematic constraints inherent to the process.   
This allows to measure all intermediate particle masses simultaneously.
Connections among those methods have also been studied 
\cite{Serna:2008zk}.

However, there are other obstacles in translating those methods into realistic applications.
The aforementioned methods, except for the inclusive $M_{T2}$ version 
\cite{Nojiri:2008hy}, 
to some extent rely on the assumption of a detailed knowledge of the
particular SUSY event (e.g.\,\,the specific production and decays).  
But in contrast, SUSY events are generally far from unique and rather 
possess a large variety of production and decay processes.
Since most SUSY events lead to similar final states with multiple jets
and missing energy, 
identification of production and decay is very difficult in a hadron collider environment.\footnote{
For an interesting study along this line, see Ref.\,\cite{Bai:2010hd}
}
The contamination from SUSY events that we are not interested in is referred to as ``SUSY background".

In general, the mass determination methods also require the knowledge
on the origin of observed particles:  which particle originates from
which decay vertex in the cascade decay chains. How much knowledge is
required depends on the corresponding method. For the edge method, the
assignments of particles which do not involve the invariant mass of
interest are irrelevant. For the inclusive $M_{T2}$ method, 
only the division of particles into two groups matters, but the
permutation inside each group is irrelevant. For the polynomial
method, the perfect particle assignment is required.\footnote{ 
Some permutations in the same decay chain may be irrelevant 
if there is a mass hierarchy between initially produced particles and the LSP
\cite{Gripaios:2011jm}.  
}
In SUSY events, gluinos and squarks promptly decay into multiple jets, leptons and the LSP. 
There is a large combinatorial number of particle assignments in the
final state to the decay chains, but any such information on the
assignment is not accessible in the detector. The wrong assignments,
called ``combinatorial background", are thus in general irreducible.     
Both, the SUSY background and the combinatorial background often cause a
serious impact on SUSY mass measurements.   

Recently, several ideas to handle the combinatorial background have been proposed 
\cite{Dutta:2011gs,Rajaraman:2010hy}.
Several studies \cite{Rajaraman:2010hy} suggest that the kinematical
edge method can effectively reduce  
the combinatorial background in the context of the $M_{T2}$-based method
in the $\tilde g \tilde g \to 4j + 2 \tilde \chi_1^0$ process.
In this method, the dijet invariant mass edge for the $\tilde g \to jj \tilde \chi_1^0$ decay is 
assumed to be already known. The position of this edge is then used as follows:
any assignment having the jet pairs exceed this gluino dijet mass edge 
is assumed to be a wrong one and rejected. Although this method
offers a good performance, the measurement (identification and
position) of the gluino dijet mass edge itself suffers from SUSY and
combinatorial backgrounds and deserves a careful study.  

The aim of this paper is to assess the feasibility of the gluino dijet
mass edge measurement in a realistic collider study including SUSY
background, the effect of initial state radiation (ISR), and a finite
detector resolution. Unlike resonance peaks,
edges (alternatively: endpoints. Invariant mass distributions of jets
from 3-body decays have rather shallow endpoints instead of pronounced
edges.) are formed by very few events only and are therefore 
intrinsically very vulnerable to any kind of backgrounds or momentum
mismeasurements. 

The relevance of ISR on the gluino mass measurement has recently been
pointed out \cite{Alwall:2009zu,Nojiri:2010mk} (for an early study on
the effect of ISR on jet measurements in LHC events,
cf.~\cite{Hagiwara:2005wg}). Events with a high $p_T$ ISR jet
intermingling with decay jets appear much more frequently within the
$\tilde g \tilde g$ than in the $\tilde q \tilde q$ production process.

In addition to ISR, we would like to emphasize the importance of the
SUSY background in the measurements. If squarks are kinematically
accessible at the LHC,  the $\tilde q  \tilde g$ associated production
has in general larger cross sections than the $\tilde g  \tilde g$ production.
The subsequent decay of the $\tilde q$ increases the number of unwanted jets and also may 
decrease the number of signal gluinos in the event if it decays to a wino or bino state directly.
Moreover, if the wino states lie in between the gluino and the LSP masses, 
gluino and left-handed squarks decay in general more frequently into
the wino states first, followed by the subsequent wino decay into the
LSP:  $\tilde g \to jj \tilde \chi_i \to jjjj \tilde \chi^0_1$ or
$\tilde q_L \to j \tilde \chi_i \to jj \tilde \chi_1^0$, where $\tilde
\chi_i$ is either $\tilde \chi_1^\pm$ or $\tilde \chi_2^0$. Those
processes do have a significant impact on the gluino dijet mass edge measurement.
The location of the edge in the $\tilde g \to jj \tilde \chi_i$
events, from hereon entitled as the ``wino edge'', is 
smaller than that in the $\tilde g \to jj \tilde \chi_1^0$ events,
which we choose to name ``bino edge''.

Therefore, in the inclusive sample, the structure of the bino edge is weakened  
because of the overwhelming $\tilde g \to jj \tilde \chi_i$ events,
and the contribution from $\tilde g \to jj \tilde \chi_1^0$ events  
overshoots the wino edge. This last point is particularly problematic,
since this overshooting is able to mimic disturbing effects of
hadronization and detector response, even when two jets from the same
decay chain and gluino jets are unambiguously selected. In this study
we attempt to disentangle those two edges by selecting events with a
particular range for the multiplicity of jets and leptons. 

Note that both the presence of high-$p_T$ ISR jets and jets from
irreducible SUSY background contributes to the combinatorial background.
For instance in a $\tilde g \tilde g \to 4j + \tilde \chi_1^0$ event, 
the ratio between correct and wrong jet pairs is $1/3$.
On the other hand, in the 
$\tilde g \tilde q + j_{\rm ISR}  \to 4j + \tilde \chi_i \tilde \chi_1^0 \to 6j + 2 \tilde \chi_1^0$ events
with two of those jets failing to satisfy a jet identification criteria (leading to the same 4$j$ + missing energy event), 
it is either $1/6$ or $0$ depending on whether or not one of the gluino jets is lost. 

In order to make the problem tractable and the study as generic as possible, 
in this paper we use a (semi-)simplified model, where
all the higgsinos, sleptons and the third generation squarks are decoupled
and an approximate grand unified theory (GUT) relation 6:2:1
is imposed on the three gaugino masses.
We also concentrate on the situation where the squarks are heavier
than the gluino for the following reasons: This type of mass spectra
is motivated by an observed excess of Higgs-like events in the
mass range around 125\,GeV, since such a relatively heavy Higgs boson
mass indicates possibly heavy third generation squarks in the MSSM.
In any case, the mass ordering between the gluino and squarks (if they
are both present at the LHC) are likely to be observed by looking at
the distributions of the hemisphere mass or the $M_{T2}$ in the high
$p_T$ dijet events~\cite{Nojiri:2010mk}.  Although the semi-simplified
model has much fewer parameters than the MSSM, it can nevertheless
cover the dominant features in most of the interesting SUSY spectra
which appear in popular SUSY-breaking scenarios such as gravity and
gauge mediation.   

The paper is organised as follows. In section 2 we introduce our
three benchmarks scenarios under investigation covering the different
kinematic and phenomenological features, as well as our event
simulation setup. In section 3, we discuss the topological event
configurations arising in our study and introduce a method of endpoint
selection by means of semi-inclusive jet mulitplicities. Then, we
propose new variables and compare them to existing ones in section 4,
give numerical results in section 5 and estimate the actual endpoint
positions in section 6 before concluding in section 7. 

 
\section{Benchmark scenarios and simulation setup}

Since our interest is the gluino three body decay, $\tilde g \to jj
\tilde B \,(\tilde W)$, we focus on scenarios with squarks heavier
than the gluino (otherwise, $\tilde g \to j \tilde q$ dominates the
gluino branching ratios). For the study of the impact of SUSY
backgrounds, we introduce a semi-simplified model in order to keep the
problem parametrically manageable and the discussion as general as possible.
The higgsino states are decoupled, therefore the lightest neutralino
is a pure composed bino state, and similarly the second lightest
neutralino and the lighter chargino are purely composed of the wino
states. We adopt an approximate GUT relation of 6:2:1 of the three
gaugino masses and fix them to $(m_{\tilde g}, m_{\tilde W}, m_{\tilde
  B}) = (1200, 400, 200)$\,GeV. The gluino dijet mass edges are then
given by $m_{jj}^{\rm max} = m_{\tilde g} - m_{\tilde B} = 1000$\,GeV
for the $\tilde g \to jj \tilde B$ process and $m_{jj}^{\rm max} =
m_{\tilde g} - m_{\tilde W} = 800$\,GeV for the $\tilde g \to jj
\tilde W$ process.        
Sleptons and third generation squarks we explicitly decouple, since
their presence may in any case help disentangle the combinatorical
issues of the underlying SUSY cascade using leptons and b-tagging (or
in the worst case not deteriorate the method presented here). 
The first two generation squarks are assumed to be degenerate and
define the following three scenarios (see also
Table\,\ref{tab:spectra}): 

\begin{description}

\item[ Scenario A  ($m_{\tilde q} = 1300$\,GeV)] ~~\\
The associated $\tilde q \tilde g$ production dominates the SUSY production processes. 
$\Delta m \equiv m_{\tilde q} - m_{\tilde g} = 100$\,GeV and the
branching fraction of $\tilde q \to j \tilde g$ is kinematically
suppressed. The squarks do mainly decay to the lighter gauginos,
$\tilde q \to j \tilde B\,(\tilde W)$, and we expect a prominently
hard jet coming from the squark decay in addition to only one signal
gluino in the dominant combined production/decay process.  

\item[ Scenario B  ($m_{\tilde q} = 1900$\,GeV)] ~~\\
The squark production starts getting suppressed because of the heavier
mass, but the associated $\tilde q \tilde g$ production still has a
sizable cross section. The mass splitting between squarks and gluino
is relatively large: $\Delta m \equiv m_{\tilde q} - m_{\tilde g} =
700$\,GeV.  The main decay mode of the squarks is $\tilde q \to j \tilde g$.
We expect a moderately hard jet coming from the squark decay as well as two signal gluinos 
in the dominant combined production/decay process.  

\item[ Scenario C  ($m_{\tilde q} = 10000$\,GeV)] ~~\\
The squarks are decoupled and not produced at the LHC.
The $\tilde g \tilde g$ process is the unique SUSY QCD production process. 

\end{description}

\begin{table}[t!]
  \centering
  \begin{tabular}{|c|c|c|c|c|c|c|}
    \hline
    spectrum & $m_{\tilde{Q}}$& $m_{\tilde{G}}$& $m_{\tilde{W}}$& $m_{\tilde{B}}$ & $m^{max}_{jj} (\tilde{W})$ &  $m^{max}_{jj} (\tilde{B})$ \\
    \hline
    A & 1300 & 1200 & 400 & 200 & 800 & 1000\\
    B & 1900 & 1200 & 400 & 200 & 800 & 1000\\
    C & 10000 & 1200 & 400 & 200& 800 & 1000 \\
    \hline
  \end{tabular}
  \caption{Relevant sparticle masses in \GeV \, for the three benchmark spectra under investigation. All other scalars and higgsinos were set to $10$ \TeV.}
  \label{tab:spectra}
\end{table}

Throughout this paper, we use the following setup of simulation tool chain: SUSY events were generated using Herwig++ \cite{Bahr:2008pv} and WHIZARD \cite{whizard}. 
Furthermore, the events are inclusive in that they are passed through the full simulation chain, i.e. decay, parton showering, hadronization and detector simulation. 
The detector responses are simulated by the DELPHES package \cite{Ovyn:2009tx} using CMS detector settings.
The anti-k$_T$ algorithm~\cite{Cacciari:2008gp,arXiv:1111.6097,hep-ph/0512210} with jet
resolution parameter $R=0.5$ is adopted, and only jets with $p_T >
50$\,GeV and $|\eta| < 2.5$ are accepted to suppress the soft
activities coming from initial and final state radiation and the underlying event.
Based on \cite{Collaboration:2011ida} the following baseline selection cuts are applied to all events:
\begin{itemize}
\item $H_T > 800$\,GeV
\item $E_{T}^{\rm miss} > 200$\,GeV
\item $\Delta \phi (j_{1/2},E_{T}^{\rm miss}) > 0.5$
\end{itemize}
where $H_T$ is defined as the scalar sum of the first four hardest jets and $j_{1/2}$ denotes the hardest or second hardest jet, respectively. These cuts are designed to suppress SM backgrounds. The cut on $\Delta \phi$ between $E_{T}^{\rm miss}$ and the hardest and second hardest jet, respectively, is applied to reject events where $E_{T}^{\rm miss}$ originates from jet energy mismeasurements.

 
\section{Selection criteria from event topologies}

\label{sec:topology}

Many existing studies address the combinatorial issue in a scenario comparable to type C and assume that the gluino has just a single decay mode: $\tilde g \to jj \tilde B$.  
However, in most of the interesting and relevant SUSY spectra which are suggested by several SUSY breaking scenarios, 
the gluino has at least two comparable decay modes:  $\tilde g \to jj \tilde B$ and $\tilde g \to jj \tilde W$, 
each of which has a different dijet mass edge.
Since the bino is lighter than the wino, the position of the bino edge
is higher than the wino edge and the $\tilde g \to jj \tilde B$
process is a serious background for the wino edge measurement. 
Because of the larger gauge coupling of the wino the gluino decays much more frequently into the wino.
Therefore even in the case that the wino edge is smaller than the bino edge, 
the $\tilde g \to jj \tilde W$ gives a significant contribution right below the bino edge which makes the bino edge measurement rather difficult.
If a gluino directly decays to a wino, the wino subsequently decays as follows:
\begin{align}
  \tilde W^0 & \rightarrow h + \tilde B \rightarrow bb \tilde B \\
  \tilde W^{\pm} & \rightarrow W^\pm + \tilde B \rightarrow jj\,(l\nu) + \tilde B 
  \label{eq:w2lep}
\end{align} 
In this study, we do not use b-tagging and treat b-jets as non-tagged jets.
As can be seen, the inclusion of the $\tilde g \to jj \tilde W$ process not only introduces a confusing extra endpoint
but also increases the number of jets in the event leading to a
drastic increase of the number of wrong jet pairings. Consequently, it
makes it more difficult to choose the correct dijet pair coming from a
gluino three body decay. In order to separetely measure two gluino
endpoints, we extract two sub-samples where one of which mostly
contains $\tilde g \to jj \tilde B$ and the fraction of $\tilde g
\to jj \tilde W$ is reasonably suppressed, and vice versa in the other
sample. To do so, we focus on the fact that the number of the final
state particles increases if the event contains the $\tilde g \to jj
\tilde W$ process. 

\begin{figure}
\centering
\begin{tabular}[t]{m{3cm}m{13cm}} 
\# decay particles & topology \\
\hline
\large{3/4} & \includegraphics[width=0.2\textwidth]{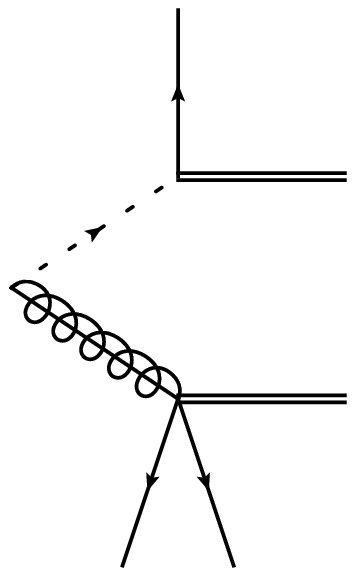}  \hspace{2mm} \includegraphics[width=0.2\textwidth]{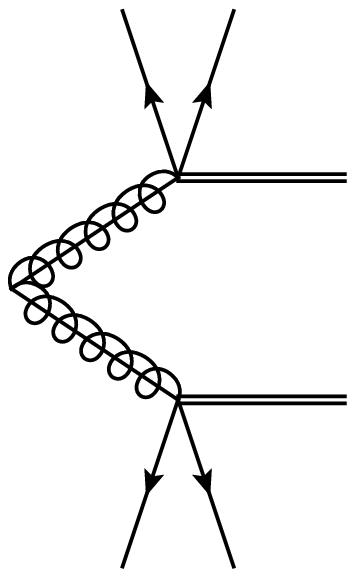}\\
\large{5} & \includegraphics[width=0.2\textwidth]{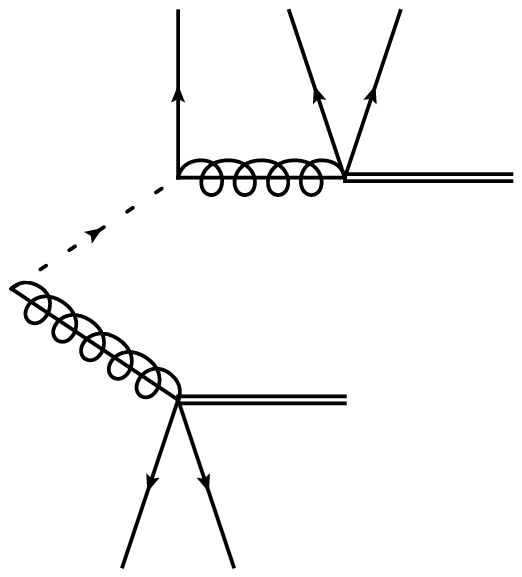} \hspace{2mm} \includegraphics[width=0.2\textwidth]{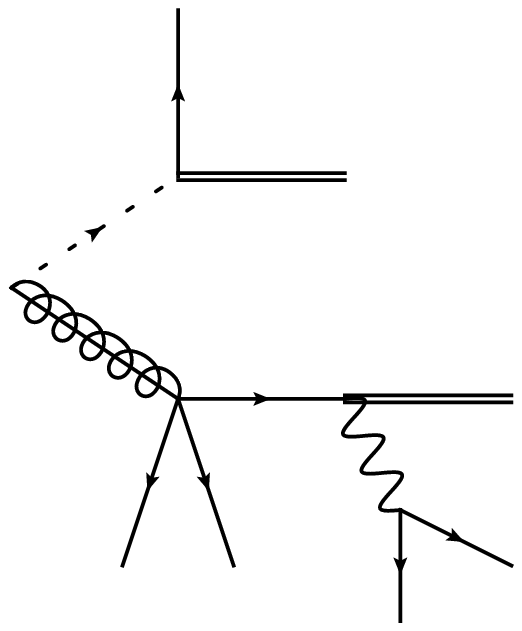}\hspace{2mm} \includegraphics[width=0.2\textwidth]{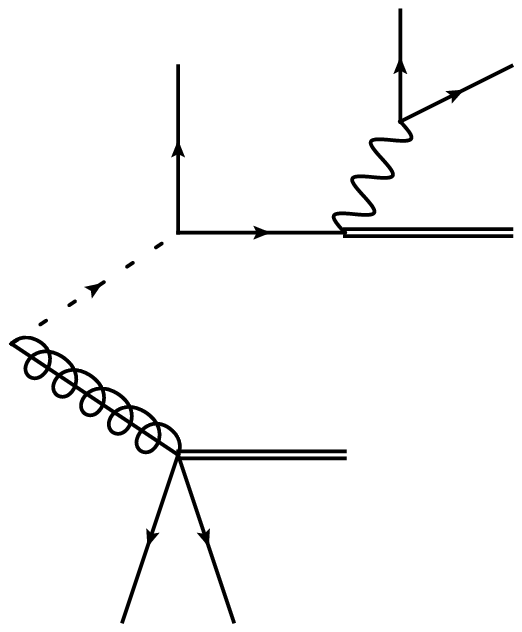}\\
\large{6} & \includegraphics[width=0.2\textwidth]{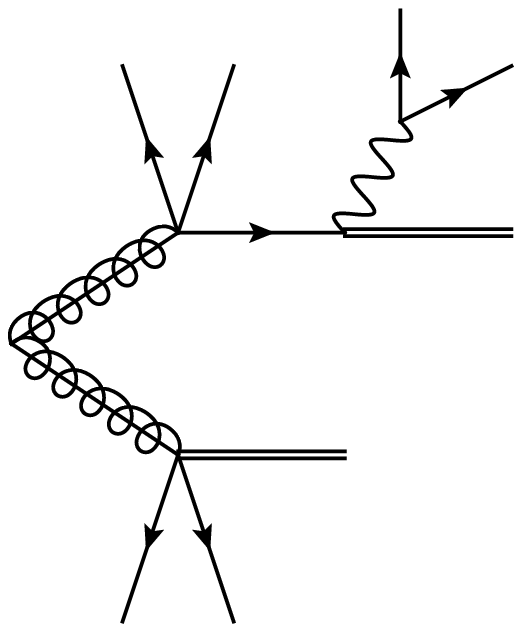} \hspace{2mm} \includegraphics[width=0.2\textwidth]{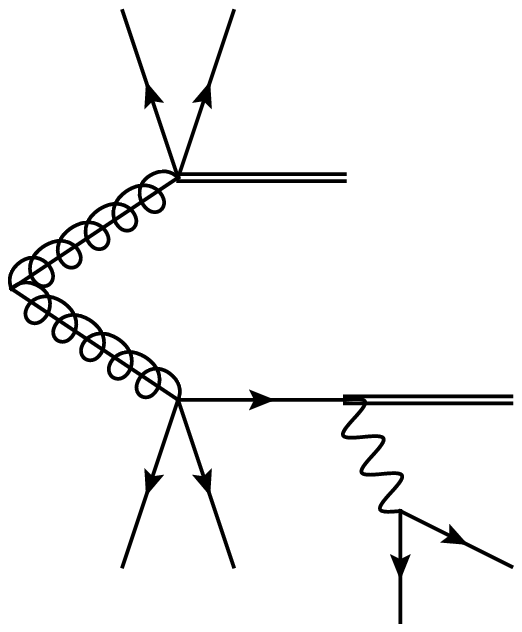}\\
\large{7} & \includegraphics[width=0.2\textwidth]{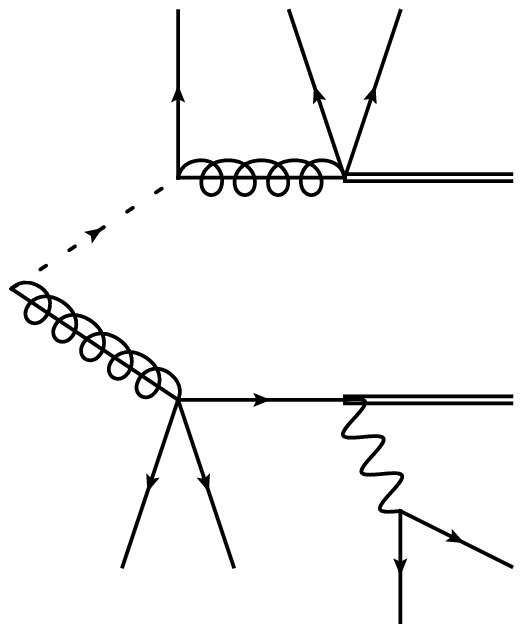} \hspace{2mm} \includegraphics[width=0.2\textwidth]{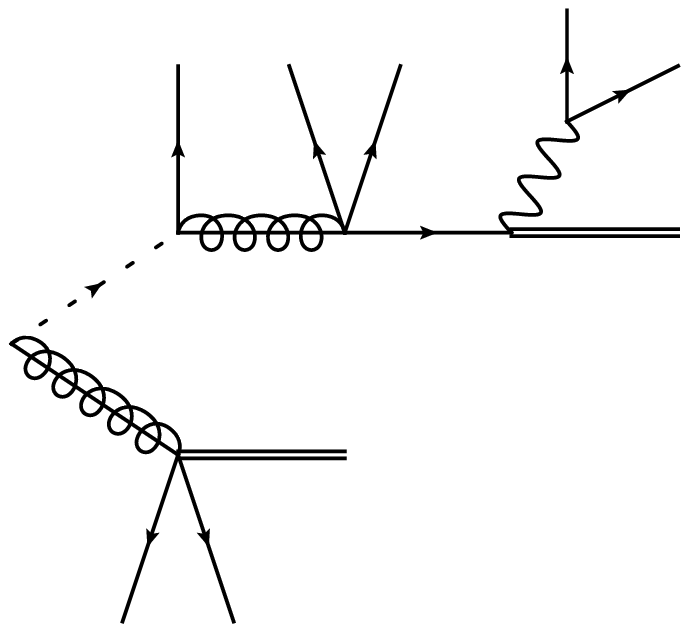}\hspace{2mm} \includegraphics[width=0.2\textwidth]{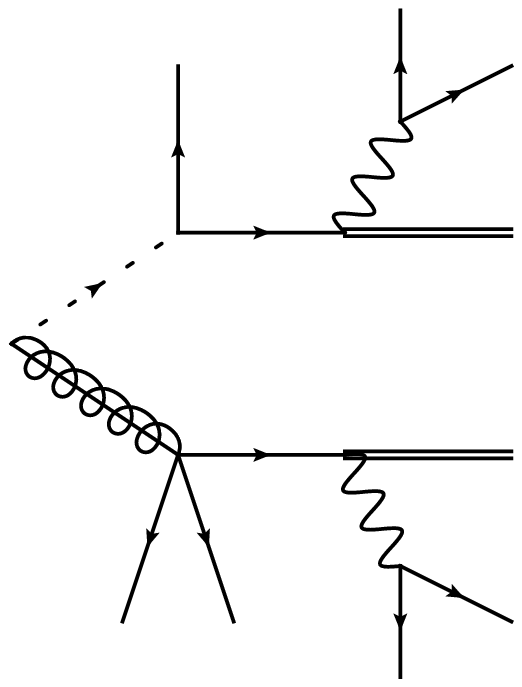}\\
\large{8/9} & \includegraphics[width=0.2\textwidth]{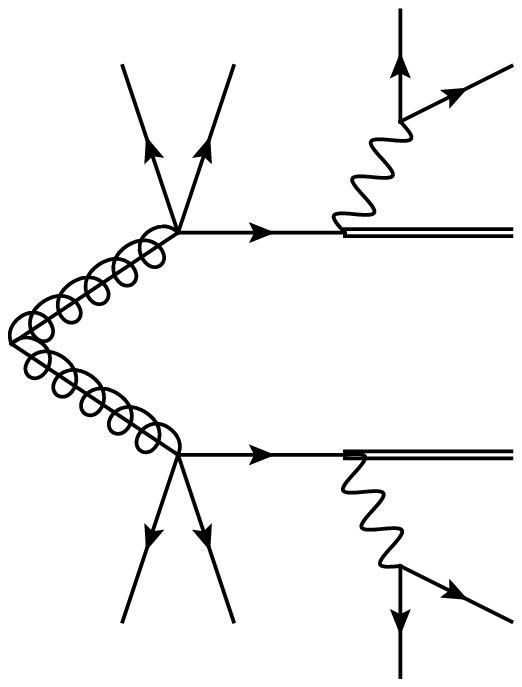} \hspace{2mm} \includegraphics[width=0.2\textwidth]{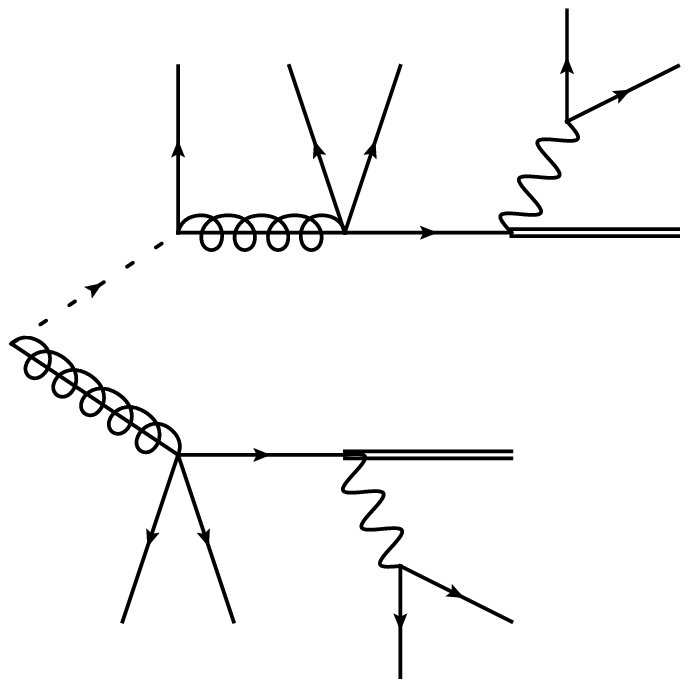}
\end{tabular}
\caption{Topological decay configurations of dominant gluino signals, sorted by number of decay particles, represented by solid lines in the final state. Double lines denote the invisible LSP.}
\label{fig:topology}
\end{figure}

In Fig.\,\ref{fig:topology}, we classify the event topology in terms of the number of decay products of SUSY cascade decay chains.
As can be seen, if we choose the events having less than or equal to four SUSY decay products, we can unambiguously select $\tilde g \to jj \tilde B$.
On the other hand, $\tilde g \to jj \tilde W$ can be unambiguously selected if we look at the events with more than or equal to eight SUSY decay products.

\begin{figure}
\centering
\includegraphics[width=0.3\textwidth]{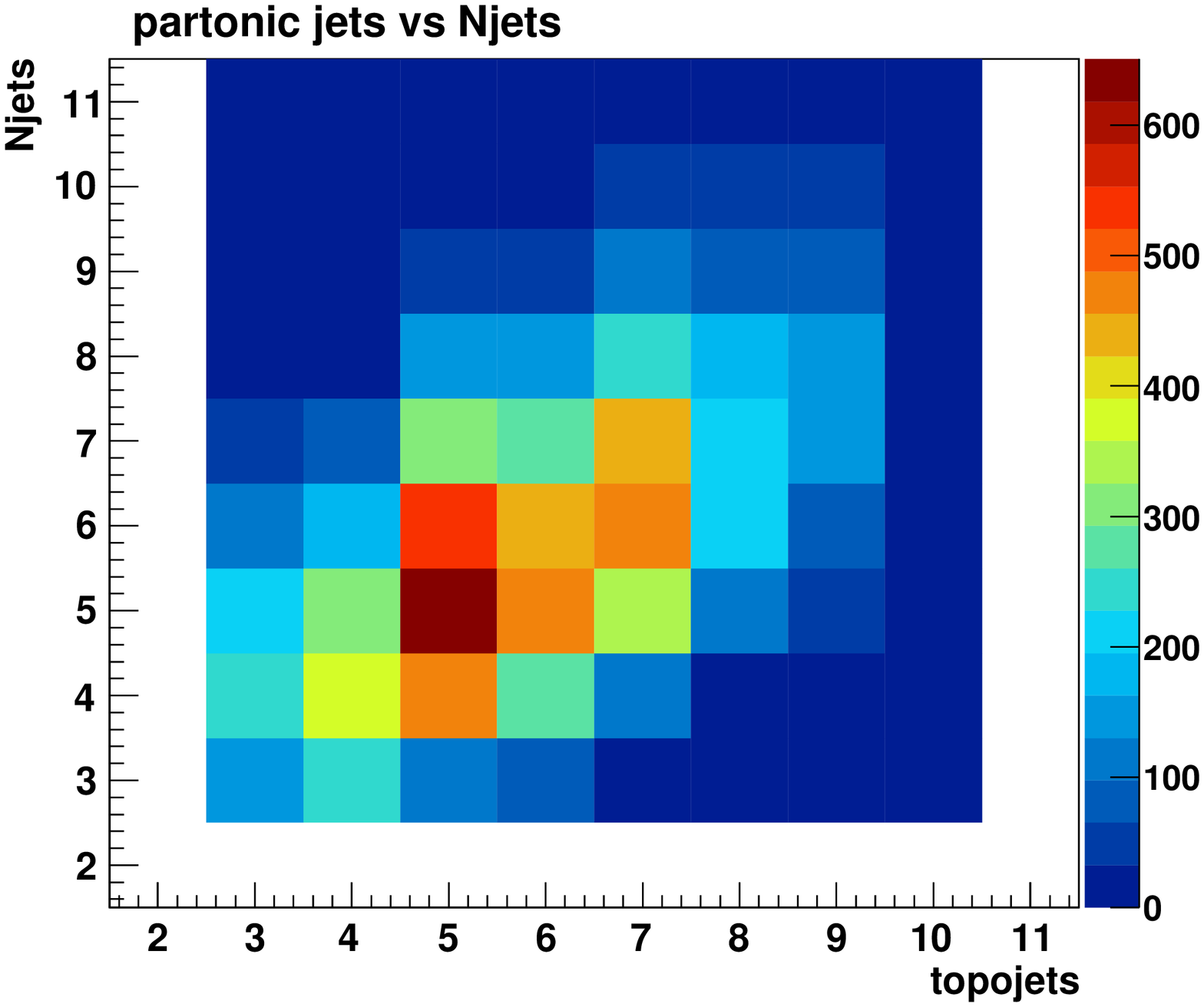}
\includegraphics[width=0.3\textwidth]{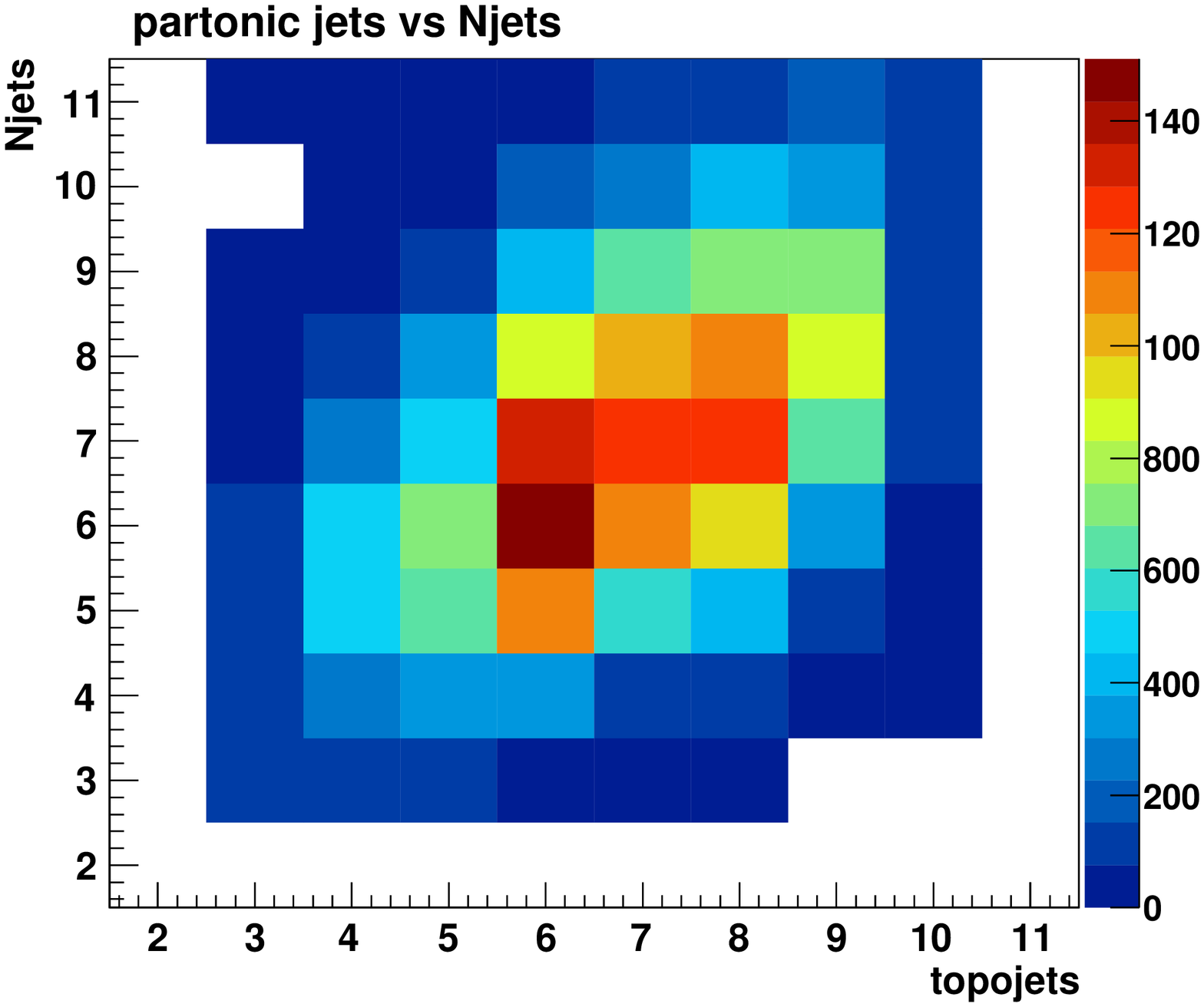}
\includegraphics[width=0.3\textwidth]{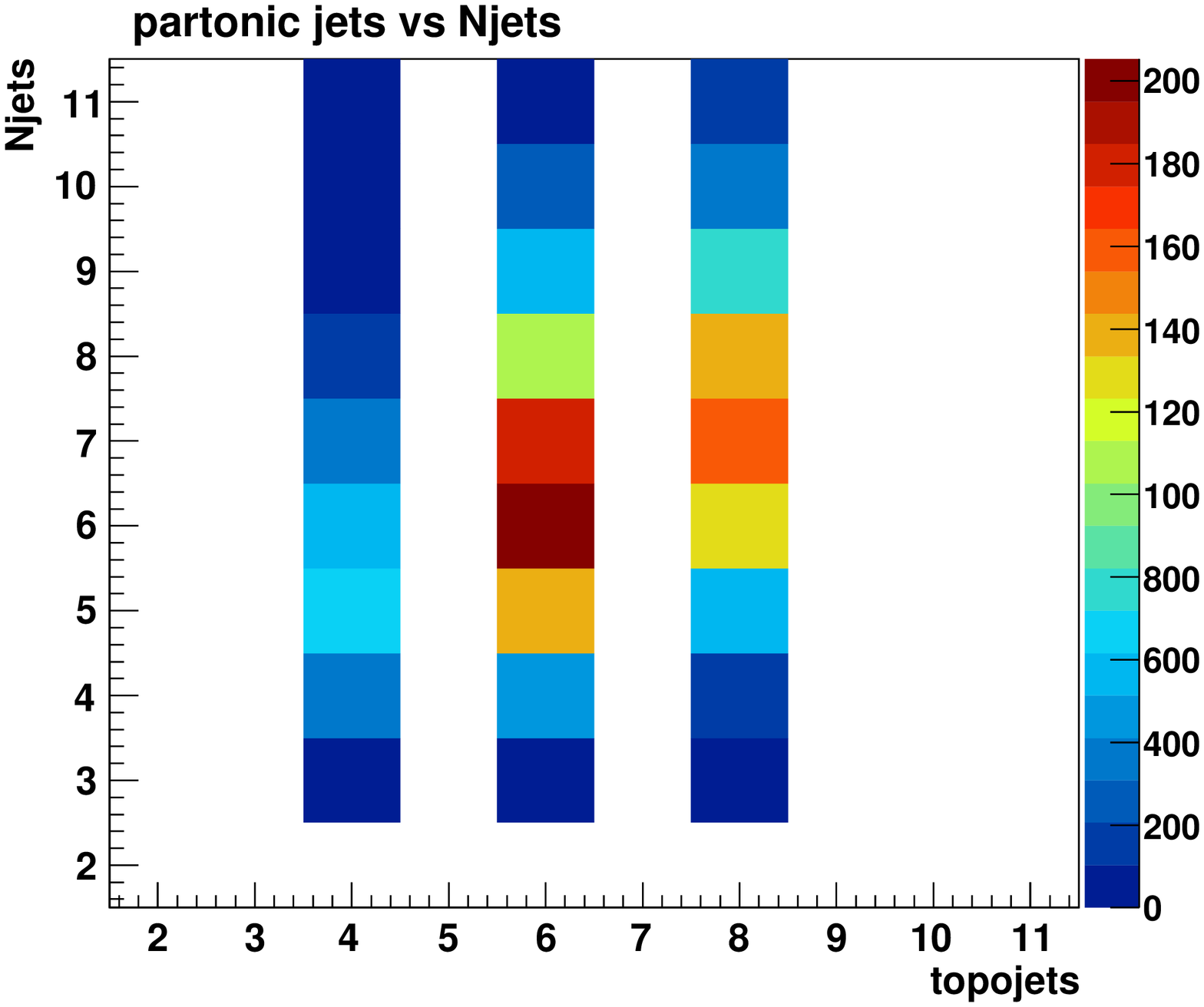}
\caption{Inclusive number of jets vs expected number of
  partonic colored particles from topological considerations. In the
  left, center and right plot scenarios of type A, B and C are shown. Cuts for detector level jets are $p_T > 50$ GeV and $|\eta| < 2.5$.} 
\label{fig:njet_topo}
\end{figure}

At detector level in a fully hadronic LHC event, it is clear that we
are not able to directly observe the number of SUSY decay products. 
However, this number should be correlated to the number of observed
jets in the detector, even after taking into account  
the effects of initial and final state radiation, underlying event,
hadronization and detector acceptances. In Fig.\,\ref{fig:njet_topo},
we show this correlation between the number of SUSY decay products and 
the number of observed jets in the detector. Keep in mind, that we only take account of jets satisfying $p_T > 50$ GeV and $|\eta| < 2.5$. As can be seen from
there, this correspondence is significantly smeared out by all the
undesired radiation, hadronization and detector effects, but it is
nevertheless still visible. This observation enables us to propose the
following two criteria for the separation of the two different edges: 

\begin{itemize}
\item \textbf{bino edge:} \quad 4-5 jets \& lepton veto,
\item \textbf{wino edge:} \quad $\ge$ 6 jets \& 1 lepton. 
\end{itemize}
These are the basis of the semi-inclusive jet multiplicity endpoint selection.
For the bino selection, we opt for 4-5 jet bins rather than selecting $\le$4 jets events.
In the 4-5 jets events, wino contamination is slightly larger than in the $\le$4 jets events.
However, the $\le$4 jets events generally suffer from rather large Standard Model backgrounds and 
a small signal cross section (see Fig.\,\ref{fig:njet_topo}). 
On the other hand, some level of wino contamination is acceptable in the bino edge measurement,
because the bino endpoint is larger than the wino endpoint, and the wino contamination is naturally suppressed 
at the vicinity of the bino endpoint. 
For the wino selection, instead of taking a $\ge$ 8 jets sample we select $\ge$ 6 jets + 1 lepton. By lepton we mean either an electron or a muon with $p_T > 20$ GeV and $|\eta|< 2.5$.
This is due to a leptonically decaying $W$, with which we are able to decrease the overall number of jets in the event by two (see Eq.\,(\ref{eq:w2lep})) and thus drastically reduce the number of wrong jet parings.

However, we should keep in mind that the correlation shown in Fig.\,\ref{fig:njet_topo} is not too strong. 
The restriction to a maximum of four or five jets thus kills a lot of
the wino contamination, but the ratio $N(\tilde{B}) / N(\tilde{W})$ of
bino to wino events still remains in the ballpark of $0.3-0.7$,
depending upon the spectrum. Extending the number of jets up to five is indispensable for types B and C since it
increases the statistics at the prize of a slightly reduced sample
purity, i.e. ratio of bino to wino events. The relative ratio of wino
to bino events in the wino selection is a lot better than in the bino
case: $N(\tilde{W}) / N(\tilde{B}) \sim \mathcal{O}(10)$.

After defining the two relevant selection criteria, we are able to
discuss the corresponding SM background contributions in more detail. 
The dominant processes after the bino endpoint selection are
QCD multijet production, where $E_{T}^{\rm miss}$ originates from
neutrinos produced in heavy flavor quark decays and jet energy mismeasurements due to instrumental effects, and the
production of neutrino pairs from $Z$ boson decays in
association with hard jets ($Z$+jets).
The production of leptonically decaying $t\bar t$ pairs
and leptonically decaying $W$ bosons in association with hard jets
($W$+jets) is expected to be suppressed by the lepton veto. 
To further reduce the background a cut on $Y_{MET}= E_{T}^{\rm miss} / \sqrt{H_T}$, which is used in SUSY searches at ATLAS and CMS \cite{Aad:2011qa,Chatrchyan:2011qs}, can be introduced in addition.

In contrast to the bino selection the wino endpoint selection criteria are expected to suppress the SM background in a way that allows to extract endpoints without applying further cuts. By requiring one lepton on top of the baseline selection mentioned above the background from QCD multijet production is anticipated to be completely suppressed. By selecting events with at least six hard jets most of $W$+jets and $Z$+jets events are also expected to be rejected.


\section{Kinematic variables for endpoint extraction}
\label{sec:var_discussion}

Our semi-inclusive jet samples contain 4 or 5 jets for
the bino selection and $\ge$ 6 for the wino selection, and 
we should therefore determine how to select gluino dijets out of a
large number of possible jet pairings. In the event samples there are
many jets that are not originating from the gluino three body decay. 
Those are mainly coming from ISR as well as from the $W$ decay.
In the scenarios A and B, an additional jet comes from the squark decay, too.
It can be argued that the gluino dijets have relatively high $p_T$
compared to the ISR and $W$ decay products, which as a kinematical
effect is easy to understand. 

\begin{figure}
\centering
\includegraphics[width=0.3\textwidth]{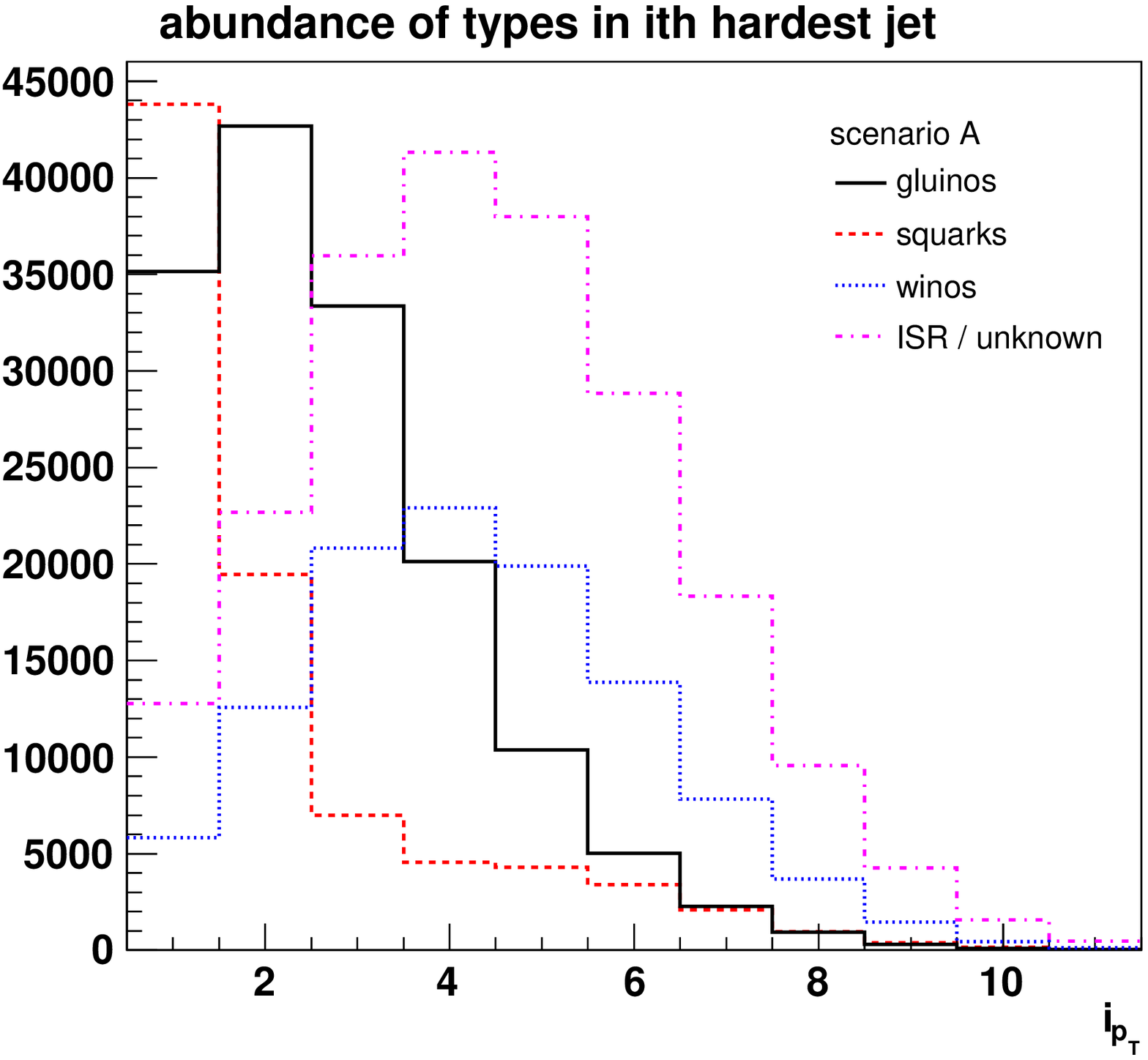}
\includegraphics[width=0.3\textwidth]{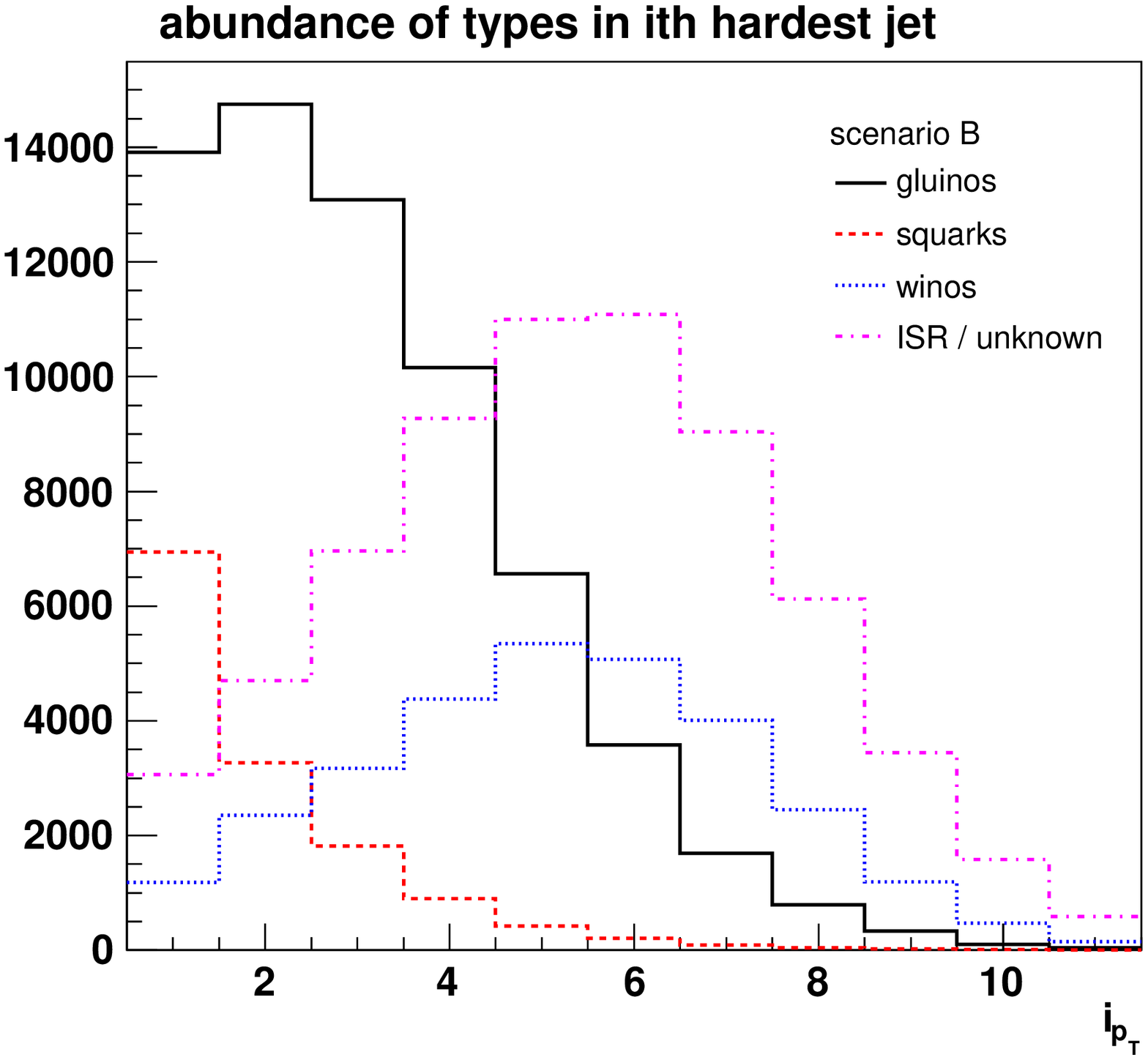}
\includegraphics[width=0.3\textwidth]{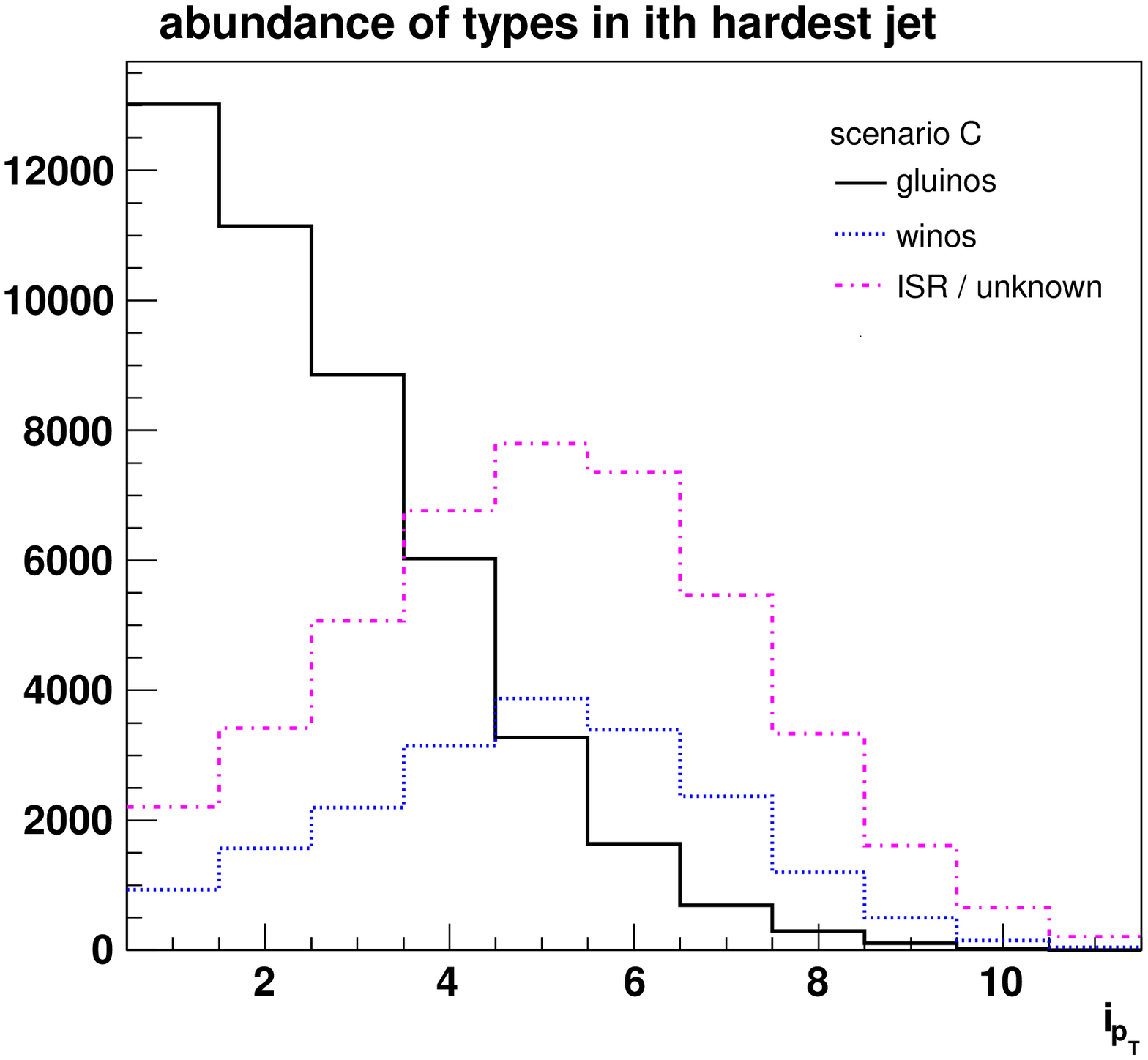}
\caption{Abundances of gluino (black full line), squark (red dashed
  line), wino (blue full line) and ISR/unknown jets (blue dotted line)
  are shown in an inclusive sample for jet bins of hardness $i$ in
  scenarios A, B and C from left to right.} 
\label{fig:ipt_abund}
\end{figure}

Figure~\ref{fig:ipt_abund} shows the abundance of gluino, squark and
wino jets in $i$-th hardest jet, sorted by $p_T$, for the three
benchmark scenarios. The identification is done using MC truth by matching detector level jets to partonic objects. Objects, which are not
successfully assigned to originate from SUSY mothers are thus either
from ISR or other clustering effects of the jet algorithm. 
As can be seen, for scenarios A, B and C, the first three 
$p_T$-hardest jets are most likely coming from the gluino decay. 
However for scenarios A and B, there is a significant contribution in the highest $p_T$ jet bin from the squark decay $\tilde q \to j \tilde B\,(\tilde W)$.
If the squark jet is wrongly selected and contributes to the distribution, 
there is a high risk to exceed the correct gluino endpoint because the dijet mass tends to be large because of the
large $p_T$ of the squark jet. This is the motivation for the minimization procedure introduced in the following variables, which takes care of this:

\begin{eqnarray}
  min_{3j} &=& \min_{k=1,2} \{ m_{3k} \}  \\
  min_{123} &=& \min_{i,j=1,2,3} \{m_{ij}\} , \qquad i \ne j
\end{eqnarray}

Here $m_{ij}$ denotes the invariant mass of jets $i$ and $j$.
The endpoint of $min_{3j}$ is expected to be the same as the gluino endpoint as long as one of the first two highest $p_T$ jets and
the third highest $p_T$ jet are coming from the same gluino decay.
The $min_{123}$ is smaller than $min_{3j}$ event-by-event because of the wider range of the minimization
and has the same endpoints as the gluino's as long as two of the first three highest $p_T$ jets are coming from the same gluino decay.      
In scenario A, the relative abundance of gluino jets does not look very promising. However
keeping in mind, that most of the time we are left with only one
gluino, the two variables proposed above we expect to also work
reasonably well.

In scenario B, it is reasonable to explicitly exclude the highest $p_T$ jet bin and build a distribution out of the remaining three hardest jets, since the gluino jets have the highest relative abundance in these bins. 
 This is the motivation behind the following variable: 
\begin{align}
  min_{234} = \min_{i,j=2,3,4} \{m_{ij}\} , \qquad i \ne j
\end{align}
Here we explicitly remove the highest $p_T$ jet and select the dijet
pair among the second, third and fourth highest jets, which yields the
smallest invariant mass. This variable should have the same endpoint
as the gluino if two of those jets are being originated from the same
gluino decay.   

There exist other methods in the literature, which address parts of
this particular problem of combinatorics in gluino endpoint
extraction. Two prominent examples are the hemisphere method
\cite{Ball:2007zza} as well as a topological method for 4 jets +
$\slashed{E}_T$ proposed in \cite{Bai:2010hd}. We give a brief
overview over both these methods as we compare them to our
kinematical variables. 

In the hemisphere method, every event is divided into two hemispheres
defined by two seeds. These are usually taken to be the hardest object
in the event and the one that maximizes the variable $\Delta R \cdot
p$, with $\Delta R = \sqrt{\Delta\eta^2 + \Delta\phi^2}$. Then all
objects are subsequently clustered to one of the two spatial areas,
defined by the seeds. This is done by assigning each particle to that
hemisphere which minimizes the value of the Lund distance measure,
\begin{equation}
  d(p_k,p^{(s)}_{j}) = (E_j-p^{(s)}_{j}\cos\theta_{jk})\frac{E_j}{(E_j + E_k)^2}
\end{equation}
between the four momentum $p_k$ of the object to be associated and
the two seed momenta $p^{(s)}_{1}$ and $p^{(s)}_{2}$. After all
objects are clustered, the seeds are updated to be the sum of all
objects in the corresponding hemisphere. Finally, the procedure is
iterated until the assignment converges (more details on the specifics
of this algorithm can be found in \cite{Ball:2007zza}). Once two
hemispheres are obtained, we pick up the first two highest $p_T$ jets
from one of the two hemispheres, which defines  
the following two variables:
\begin{equation}
m^{(1)}_{12} = m_{12}~~({\rm from~hemisphere~1}),~~~~m^{(2)}_{12} = m_{12}~~(\rm from~hemisphere~2),
\end{equation}  
where hemisphere 1 is defined as the hemisphere which contains the highest $p_T$ jet in the event.\\

Concerning the second method, in ref.~\cite{Bai:2010hd}, the authors
have studied the possibility of identifying the dominant event
topology in exclusive 4 jet + $\slashed{E}_T$ events.   
For this purpose, they defined two dijet mass variables, $F_3$ and $F_4$.
$F_3$ is designed for event topologies where 3 jets come from the same
cascade chain and 1 jet from the other one.  It is given by
\begin{eqnarray}
  F_3(p_1,p_2,p_3,p_4) = m_{kl}, \qquad \text{ for} \quad
  \epsilon_{ijkl} \ne 0 \quad\text{and} \quad m_{ij} =
  \max_{r,s=1,\dots,4}\{ m_{rs} \} 
\end{eqnarray}
which is the invariant dijet mass \textit{opposite} to the maximum of all possible dijet masses.
This variable has the same endpoint as the largest dijet mass endpoint
originating from the cascade chain producing the 3 jets.  
$F_4$ is, in contrast, designed for the symmetric event topology where
both the cascade chains produce 2 jets each.
The definition of this variable is given by
\begin{equation}
  F_4(p_1,p_2,p_3,p_4) = \min_{i,j=1,\dots,4}\{ 
    \max \left(m_{ij},m_{kl}\right) \}, \qquad
  \epsilon_{ijkl} \ne 0. 
\end{equation}
which is the minimum of the larger dijet mass pair out of three possible combinations.
It has the same endpoint as the dijets coming from the same cascade chain each.
Although those variables have originally been defined to address
exclusive 4 jet + $\slashed{E}_T$ events, we use them for our bino
and wino selection samples by applying them to just the first four highest $p_T$
jets. 


\section{Disentangling gluino endpoints}

In this section, we show the distributions of the variables defined in
the previous section. 
We assume LHC  at 14\,TeV with an integrated luminosity of 300
fb$\mbox{}^{-1}$, which corresponds to 108k, 27.6k, and 16.2k
signal events for scenarios of type A, B and C, respectively, according to
the leading order cross section computed by Herwig++.  Notice, that
these numbers are conservative in that SUSY QCD NLO corrections for
squark and gluino production are known to increase cross sections by a
K-factor of up to 2 \cite{Beenakker:1996ch}.  In the simulation we
take account of all the QCD productions, possible decays, and the
effects of parton showering, hadronization and detector simulation.   

\subsection{Scenario A}

\begin{figure}[t!]
\centering
\textbf{bino} \hspace{6.5cm} \textbf{wino}\\
\includegraphics[width=0.45\textwidth]{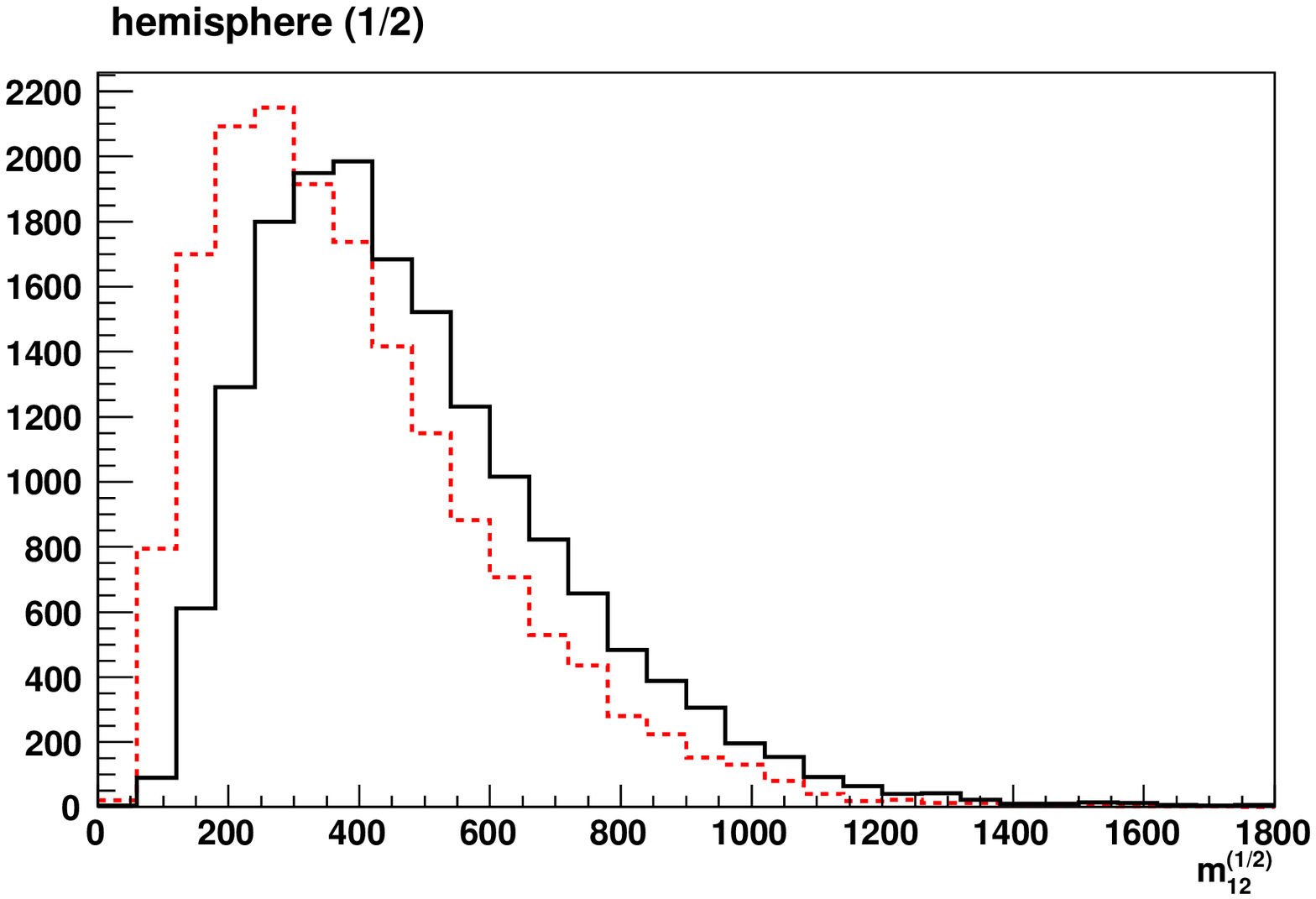}
\includegraphics[width=0.45\textwidth]{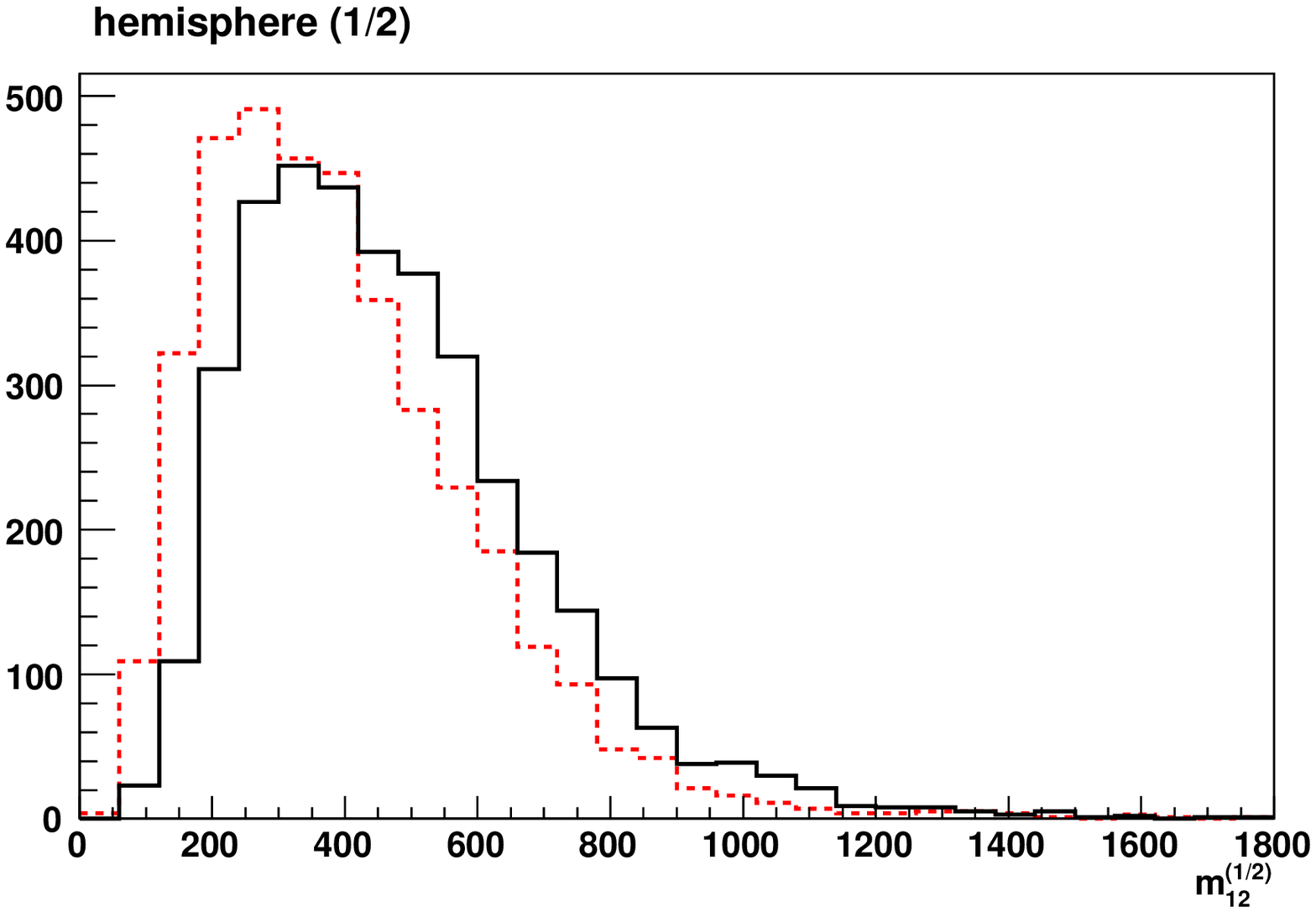}\\
\includegraphics[width=0.45\textwidth]{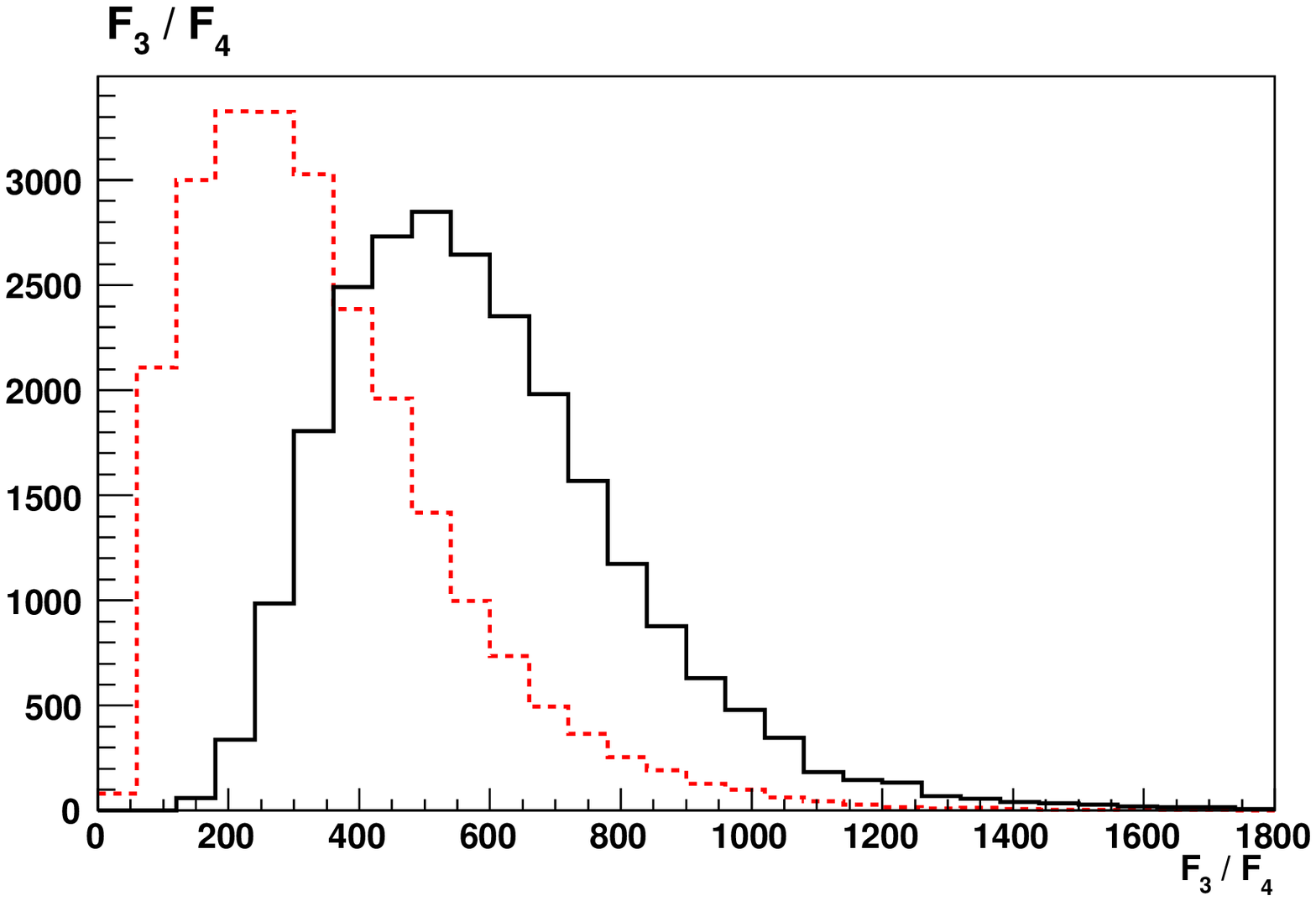}
\includegraphics[width=0.45\textwidth]{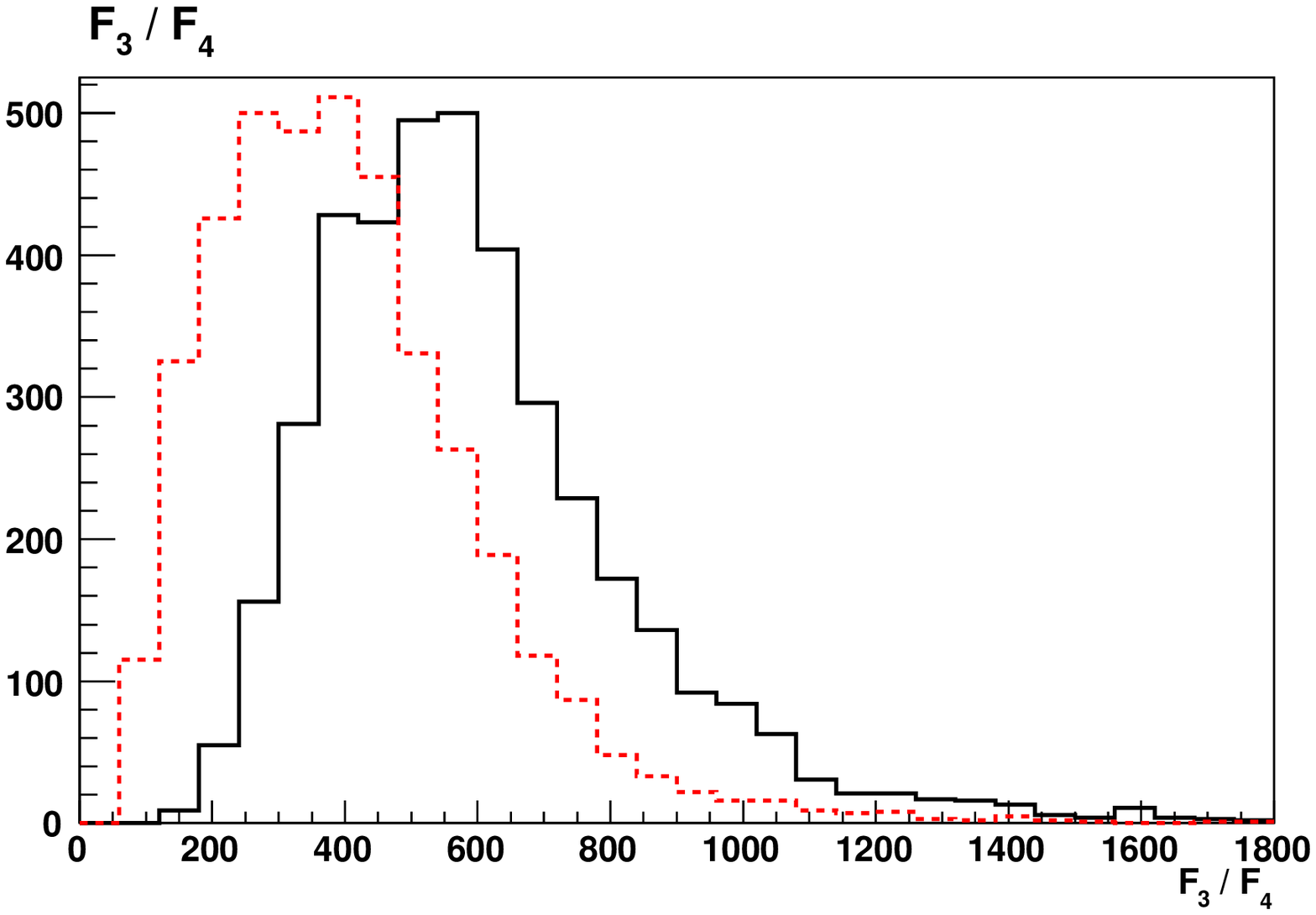}\\
\includegraphics[width=0.45\textwidth]{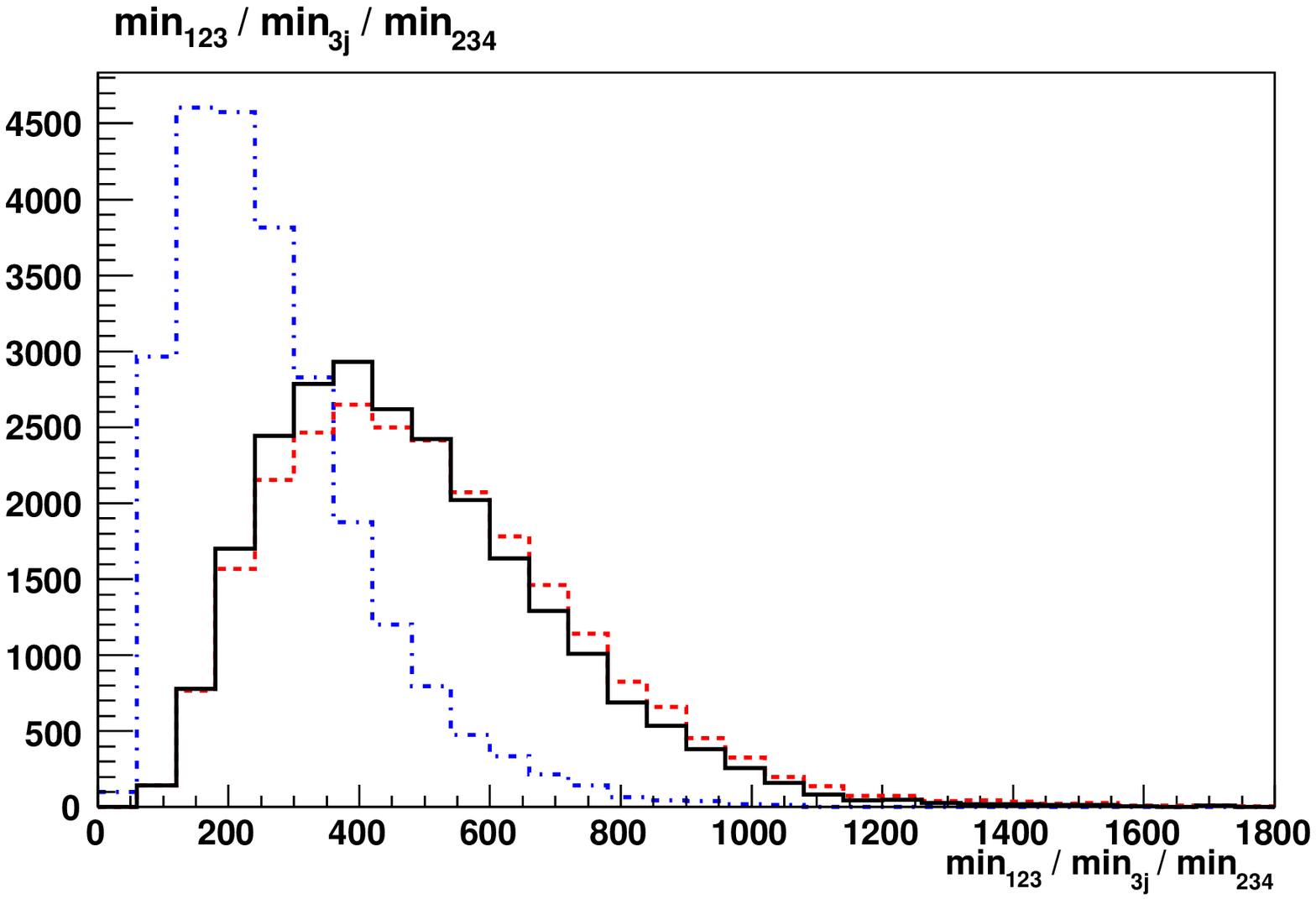}
\includegraphics[width=0.45\textwidth]{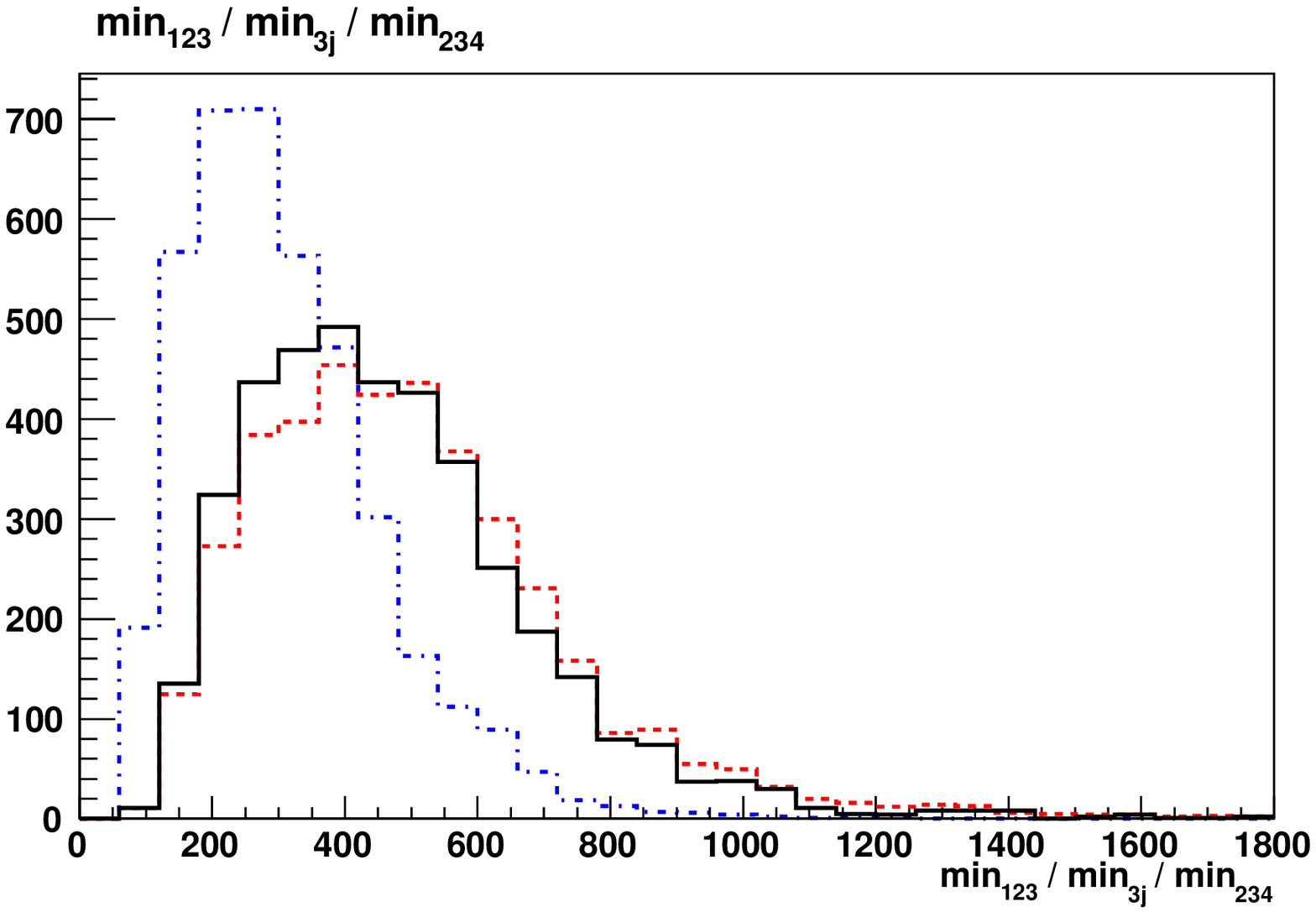}
\caption{Scenario A: first row: invariant mass of two hardest objects $m_{12}^{(1)}$ and $m_{12}^{(2)}$ for hemispheres 1 (solid, black) and 2 (dashed,red). Second row: $F4$ (solid, black) and $F3$ (dashed,red). Third Row: $min_{123}$ (solid, black), $min_{3j}$ (dashed,red) and $min_{234}$ (dot-dashed, blue). All variables are defined in section \ref{sec:var_discussion}. The left and right column correspond to the bino and wino selection criteria, respectively.}
\label{fig:mjet_typeA}
\end{figure}

In scenario A, the mass splitting between squark and gluino is only 100\,GeV.
The associated $\tilde q \tilde g$ process dominates the SUSY production.
The squarks decay preferably into bino or wino directly because the
decay mode into gluino is kinematically suppressed. 
This reduces the number of signal gluinos from two to one compared to the other scenarios.
The `squark jet' coming from $\tilde q \to j \tilde B\,(\tilde W)$ has
a significant $p_T$ (c.f. Fig.\,\ref{fig:ipt_abund}) 
and it should be avoided in favorable dijet combinations, either by
minimization or explicit removal for example in the construction of
$min_{234}$.   
 
Fig.\,\ref{fig:mjet_typeA} shows the distributions of the various variables
discussed in the previous section for the scenario A. In the left
(right) row, the bino (wino) event selection (4 or 5 jets for bino,
$\ge$ 6 jets \& $\ge$ 1 lepton for wino) is applied. In the two plots
on the top, the dijet is chosen as the first two highest $p_T$ jets in
one of the two hemisphere groups. The black solid histograms consider
the hemisphere 1, which contains the highest $p_T$ in the event and
the red dashed histograms consider the other hemisphere group
(hemisphere 2). In the middle line, the black solid histograms show
$F_4$ and the red dashed represent $F_3$. The bottom plots
show $min_{123}$ (black solid), $min_{3j}$ (red dashed) and $min_{234}$ (blue dot-dashed). 

For the bino edge measurement, $F_3$ and $F_4$ fail to recover the correct edges.
$min_{123}$ and $min_{3j}$ have very similar distributions, but nonetheless both have a slight tendency to overshoot the correct endpoint.
The hemisphere variables $m_{12}^{(1)}$ and $m_{12}^{(2)}$ also have endpoint structures around the true bino edge. Especially $m_{12}^{(2)}$ from the softer hemisphere looks most promising out of the examined variables. This is expected due to typically only one gluino and the asymmetric nature of the signal in scenario A.
  
For the wino edge measurement, most of the distributions have tails above the correct value. These tails are bigger for $min_{123}$ and $min_{3j}$ and a kink structure is less pronounced.
However, both $m_{12}^{(2)}$ and $F_3$ show a nice edge structure at the vicinity of the true endpoint.

\subsection{Scenario B}

\begin{figure}[t!]
\centering
\textbf{bino} \hspace{6.5cm} \textbf{wino}\\
\includegraphics[width=0.45\textwidth]{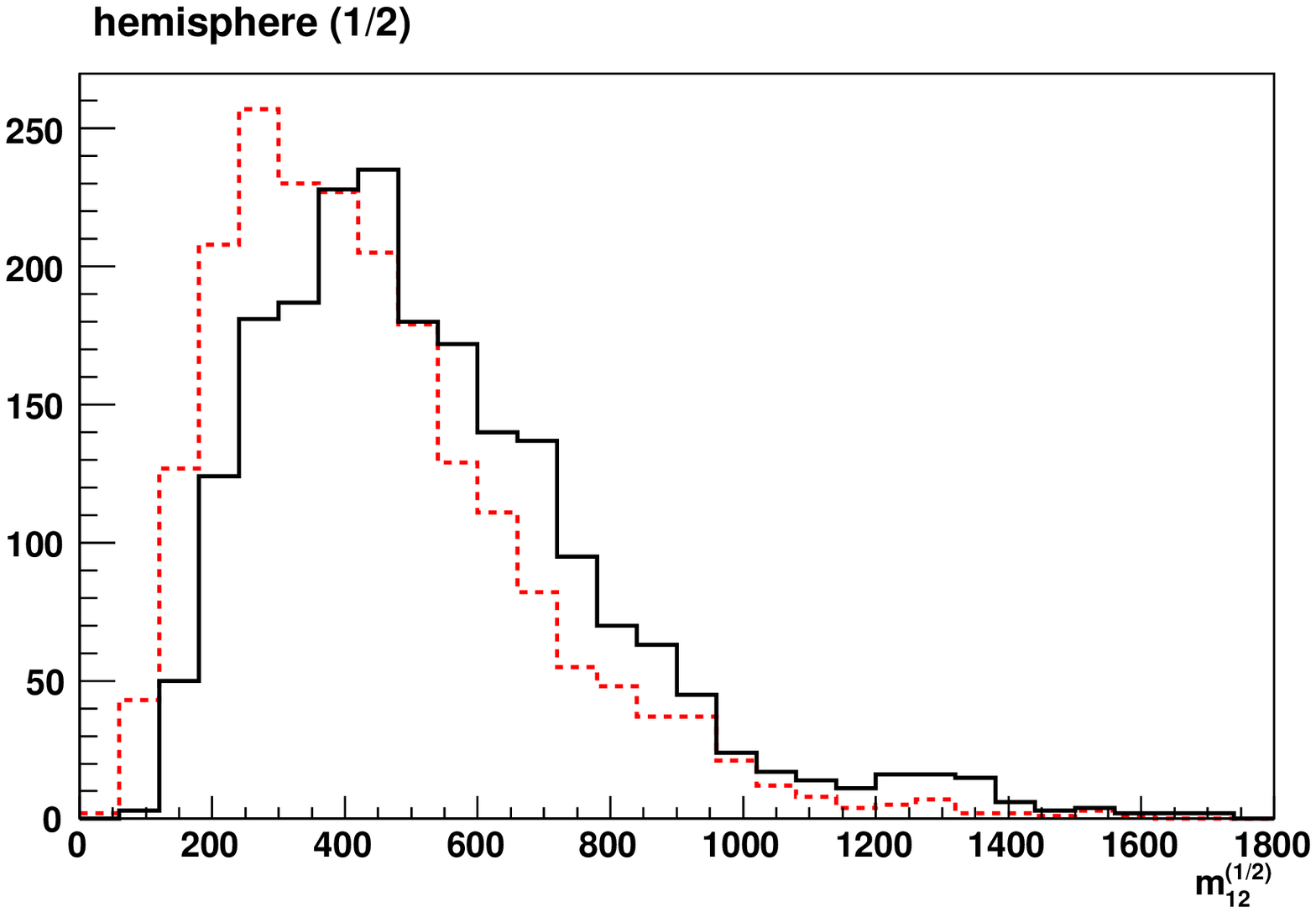}
\includegraphics[width=0.45\textwidth]{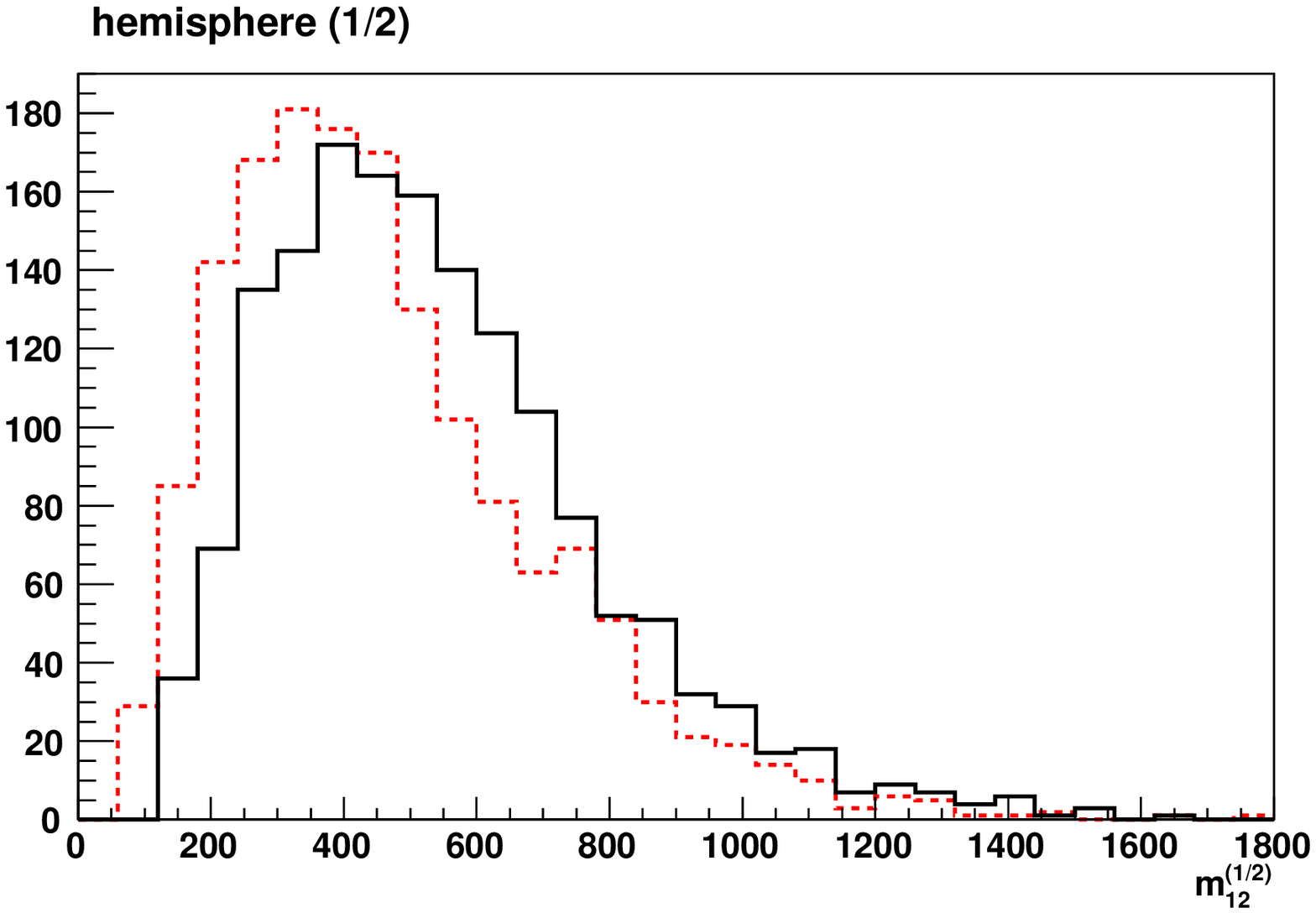}\\
\includegraphics[width=0.45\textwidth]{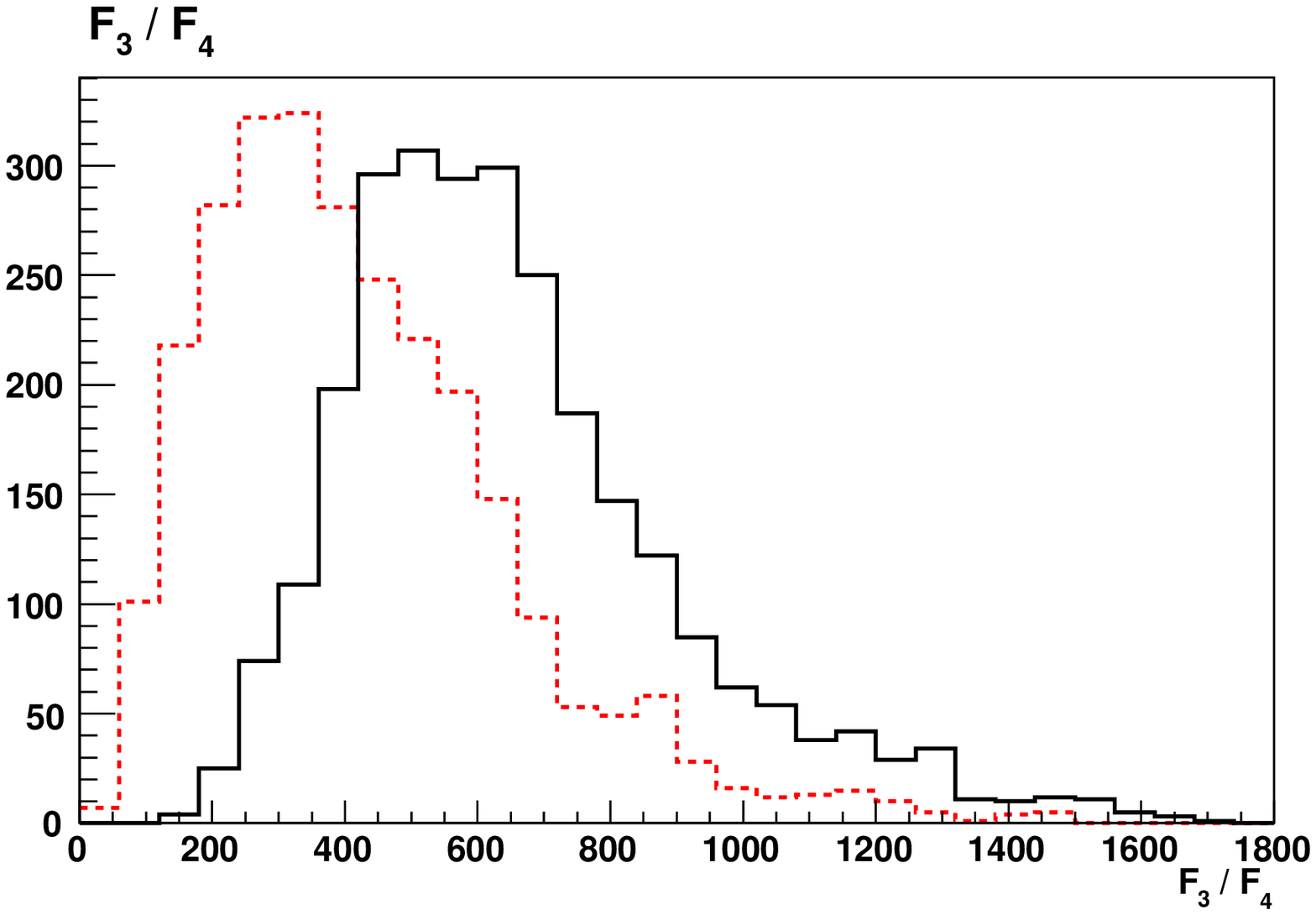}
\includegraphics[width=0.45\textwidth]{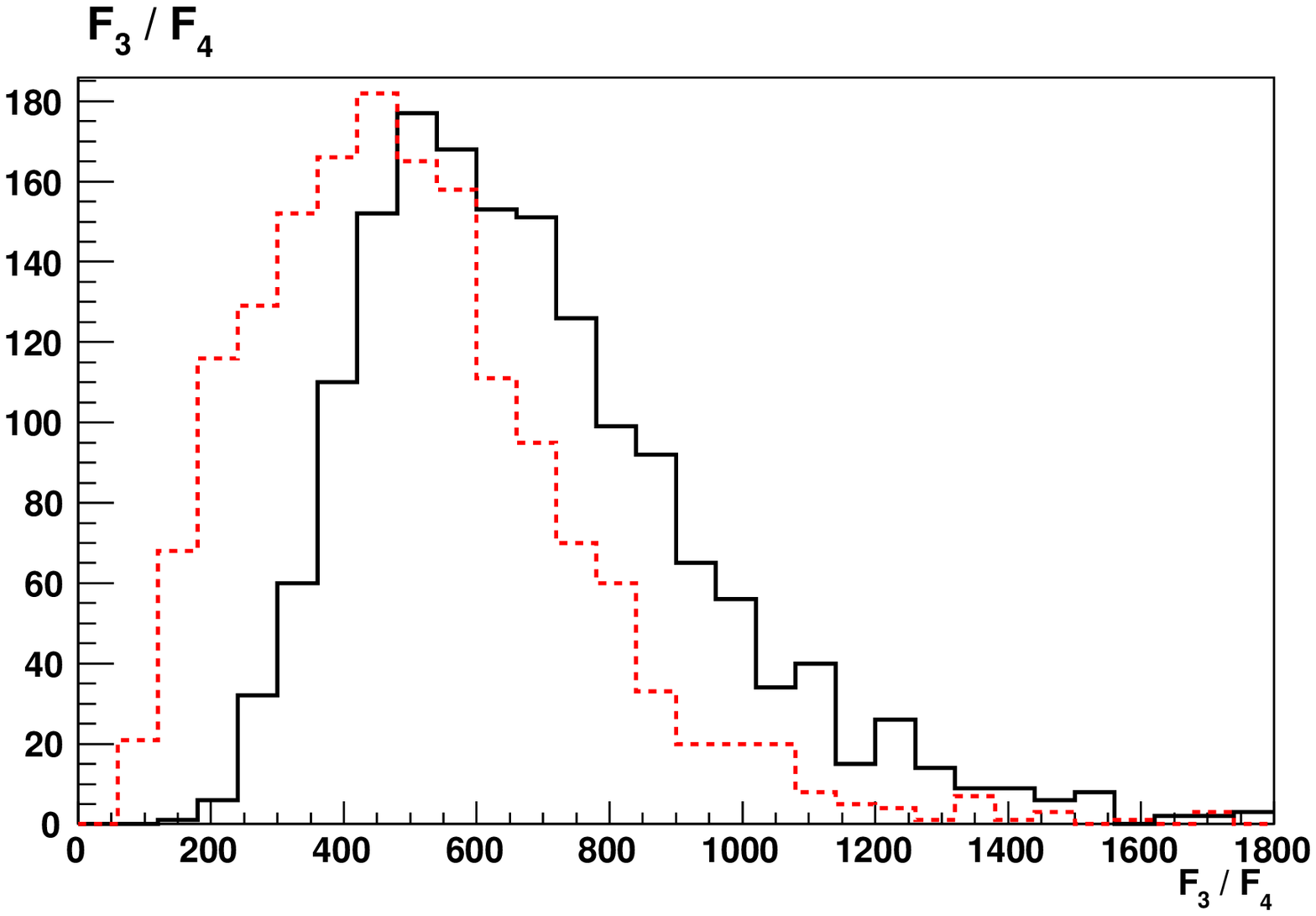}\\
\includegraphics[width=0.45\textwidth]{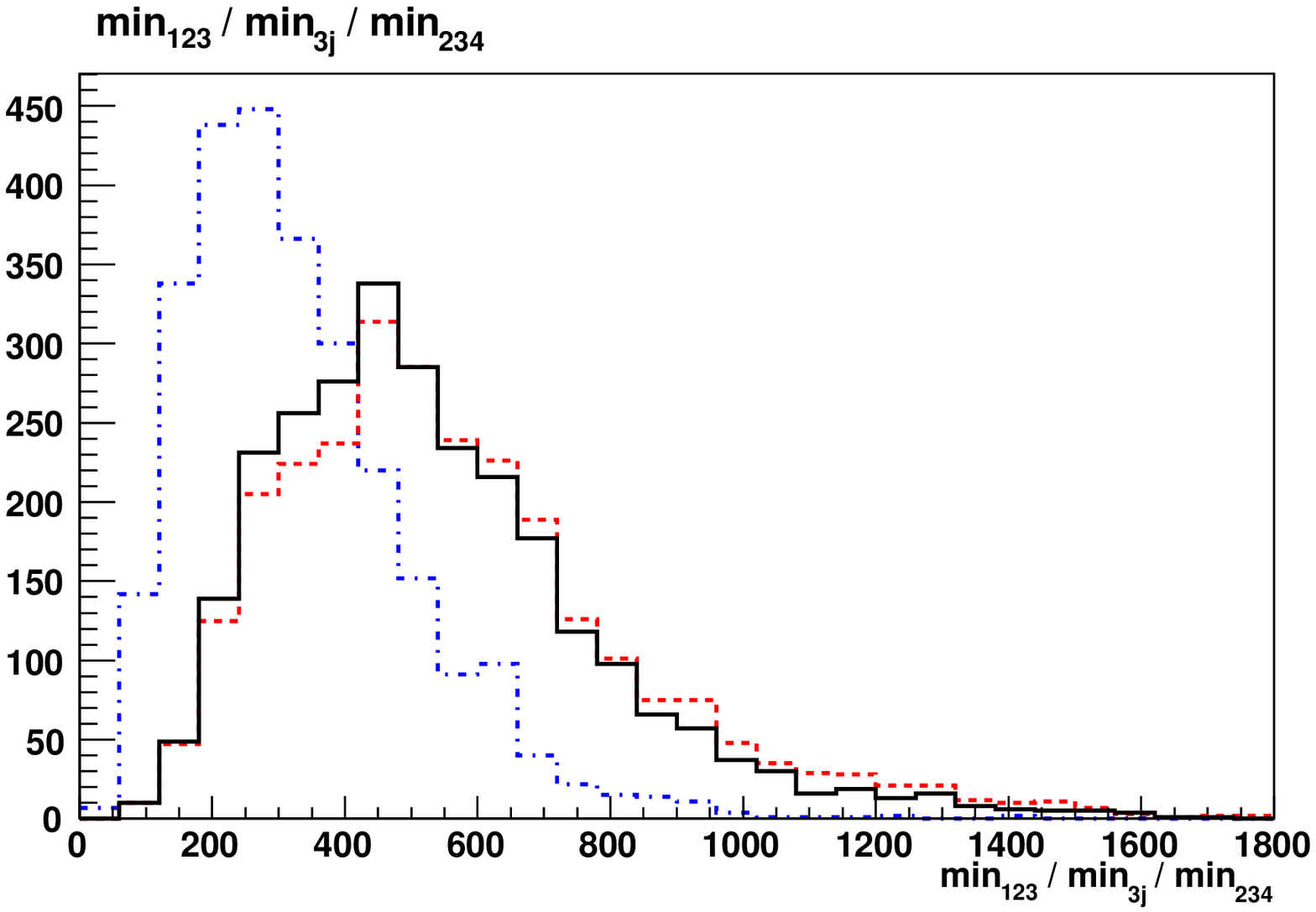}
\includegraphics[width=0.45\textwidth]{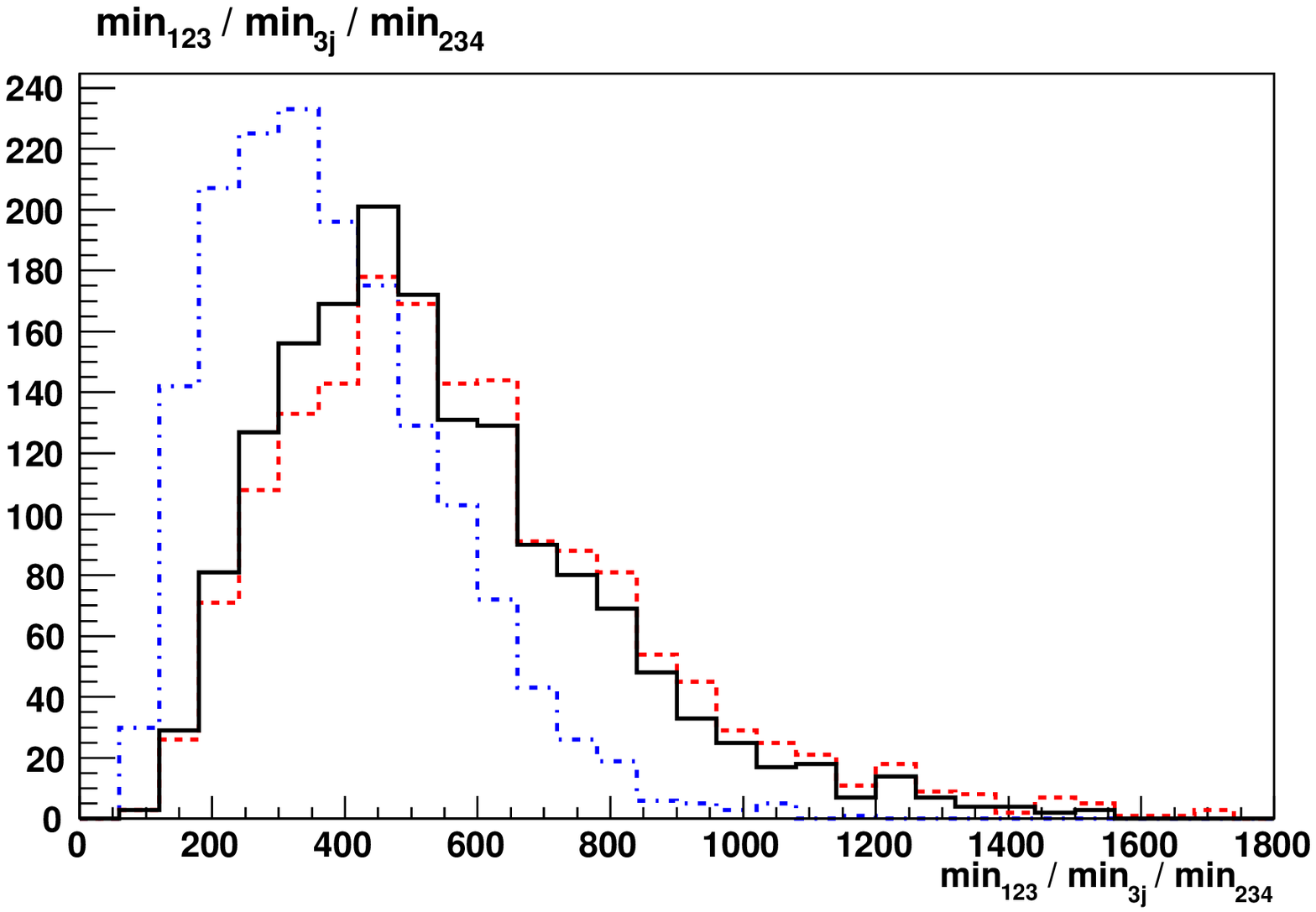}
\caption{Scenario B: performance of the variables under investigation. For details, see description of Fig.\ref{fig:mjet_typeA}.}
\label{fig:mjet_typeB}
\end{figure}

In the type B spectrum, the squark is marginally heavier than in
scenario A and the associated $\tilde q \tilde g$ production still has
a sizable cross  section compared to the $\tilde g \tilde g$ process.
The main squark decay mode is $\tilde q \to j \tilde g$. As this
`squark jet' has a large $p_T$ compared to the other jets 
(c.f. Fig.\,\ref{fig:ipt_abund}), it is quite problematic because it
is likely to be in the first three highest $p_T$ jets and thus 
increases the number of wrong jet pairings.

Fig.\,\ref{fig:mjet_typeB} shows the distributions of the variables in
scenario B. It is obvious that many of the distributions get shifted
to higher mass regions and overshoot the true endpoint. For the bino
edge measurement, the $m_{12}^{(1)}$ and $m_{12}^{(2)}$ variables
obtained from the hemisphere method work quite well.
The $F_4$ variable significantly overshoots the true endpoint, while
$F_3$ seems to recover the gluino endpoint, although the slope around
the endpoint is quite shallow. The distributions of $min_{123}$ and
$min_{3j}$ are again very similar.  They exhibit a long tail above the gluino
mass edge but some structure seems still visible at the vicinity of
the true endpoint. 

For the wino edge measurement, most of the variables fail to recover the correct endpoint.
$min_{234}$ on the other hand has a structure around the true wino edge.
A good behaviour for $min_{234}$ is expected in scenario B, since the
highest $p_T$ jet bin has a dangerous contamination from the squark
jet, and gluino jets are more safely to be picked up from the second,
third and fourth highest $p_T$ jets (cf. Fig. \ref{fig:ipt_abund} or section \ref{sec:var_discussion}).  

\subsection{Scenario C}

\begin{figure}[!t]
\centering
\textbf{bino} \hspace{6.5cm} \textbf{wino}\\
\includegraphics[width=0.45\textwidth]{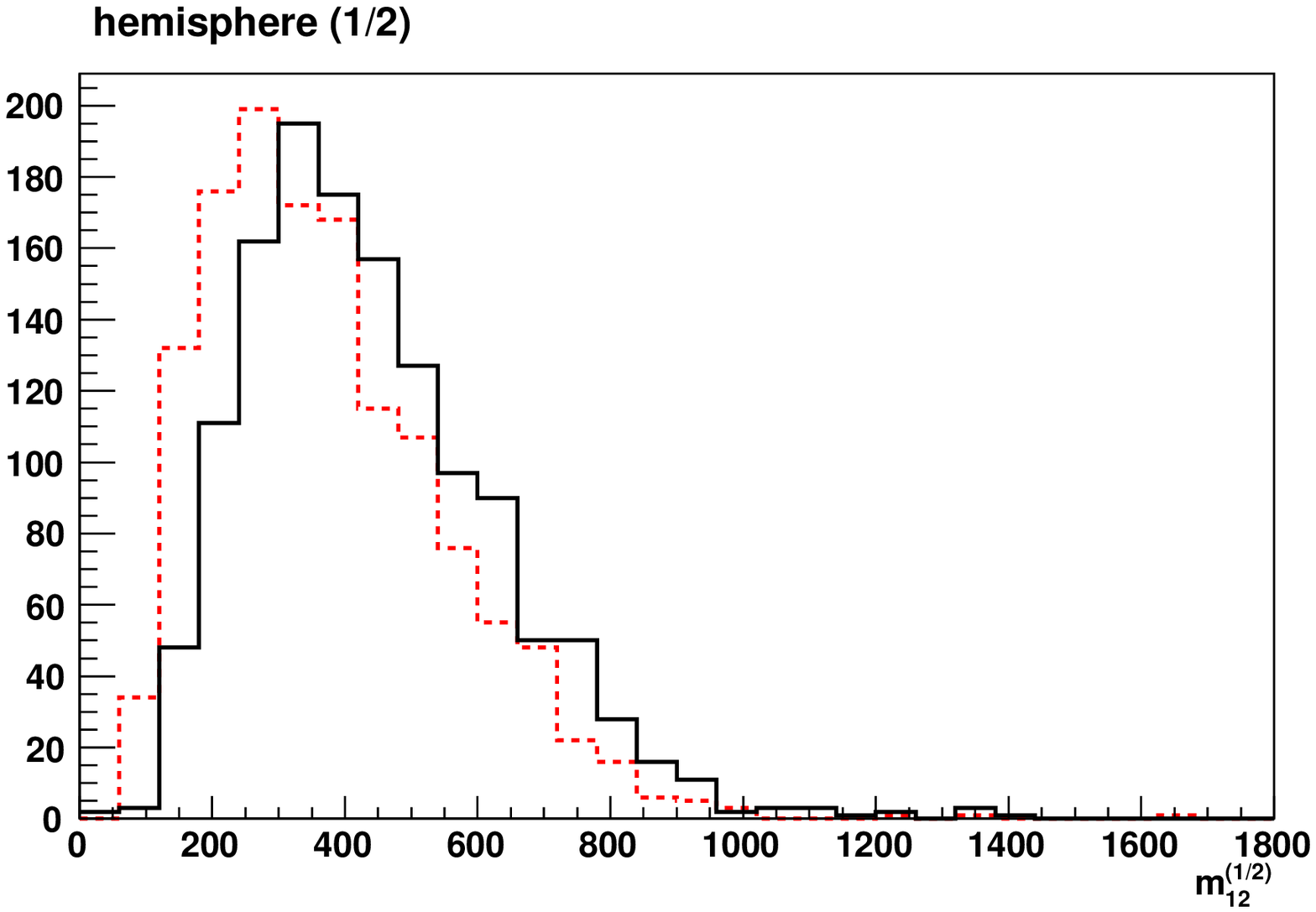}
\includegraphics[width=0.45\textwidth]{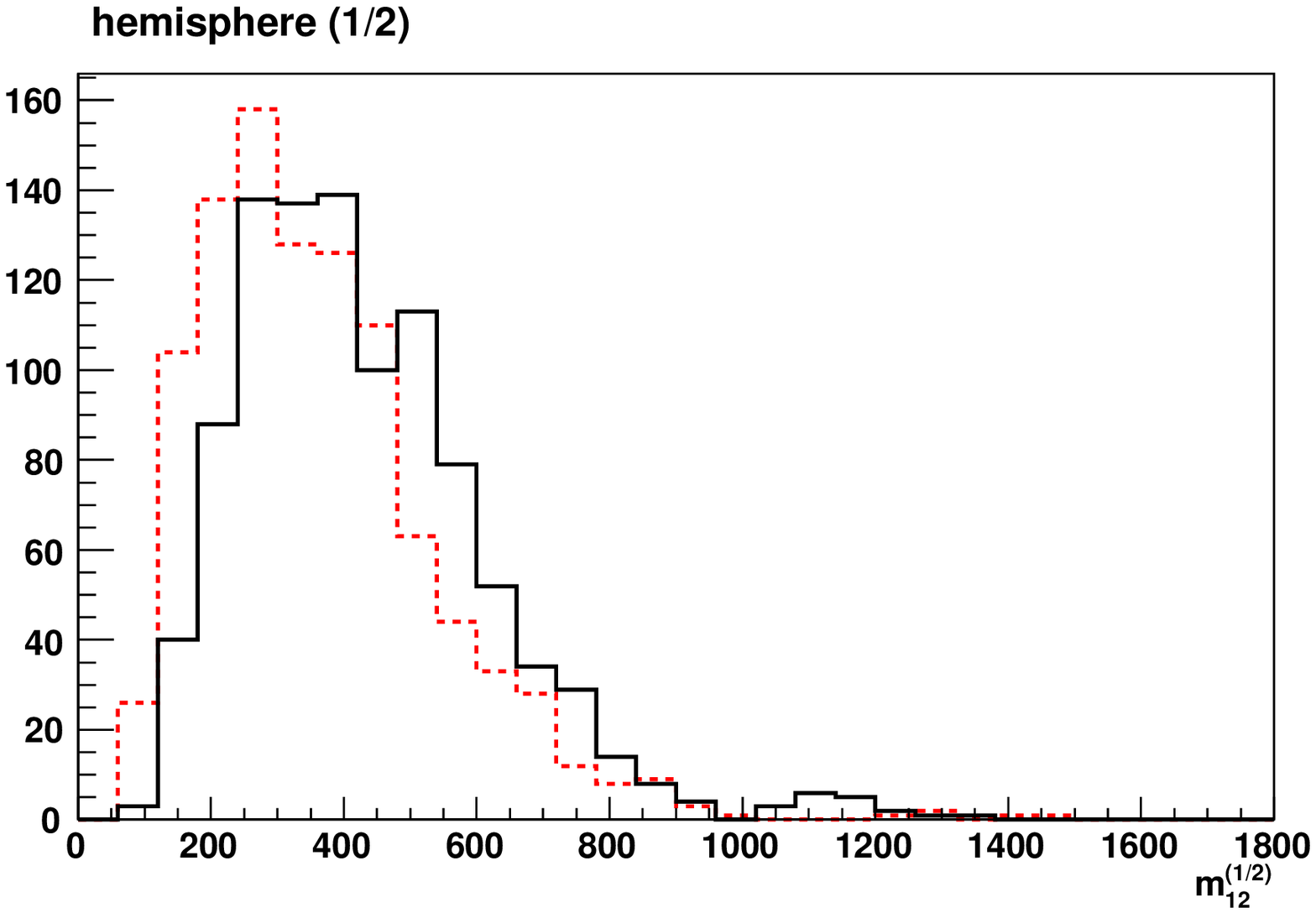}\\
\includegraphics[width=0.45\textwidth]{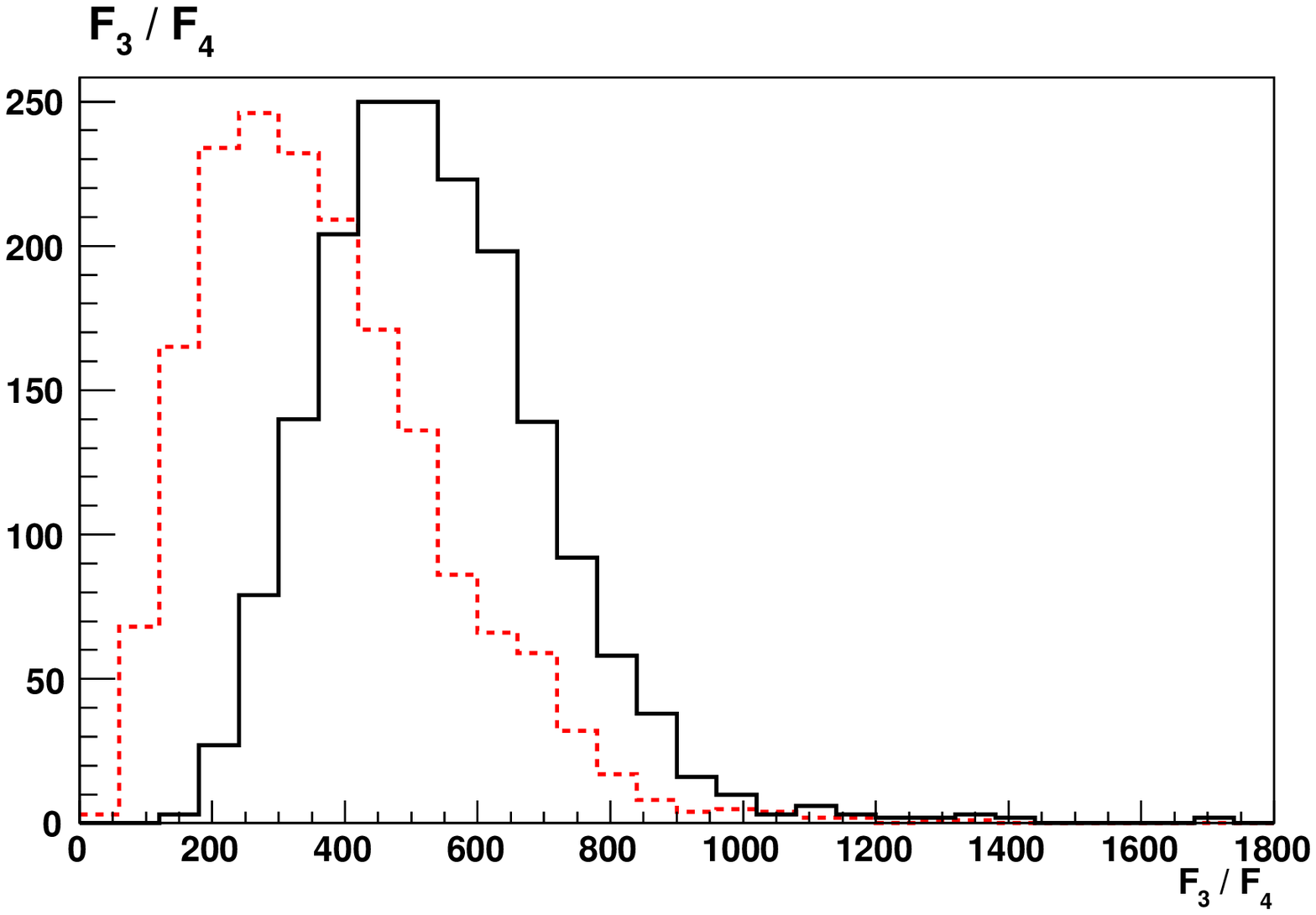}
\includegraphics[width=0.45\textwidth]{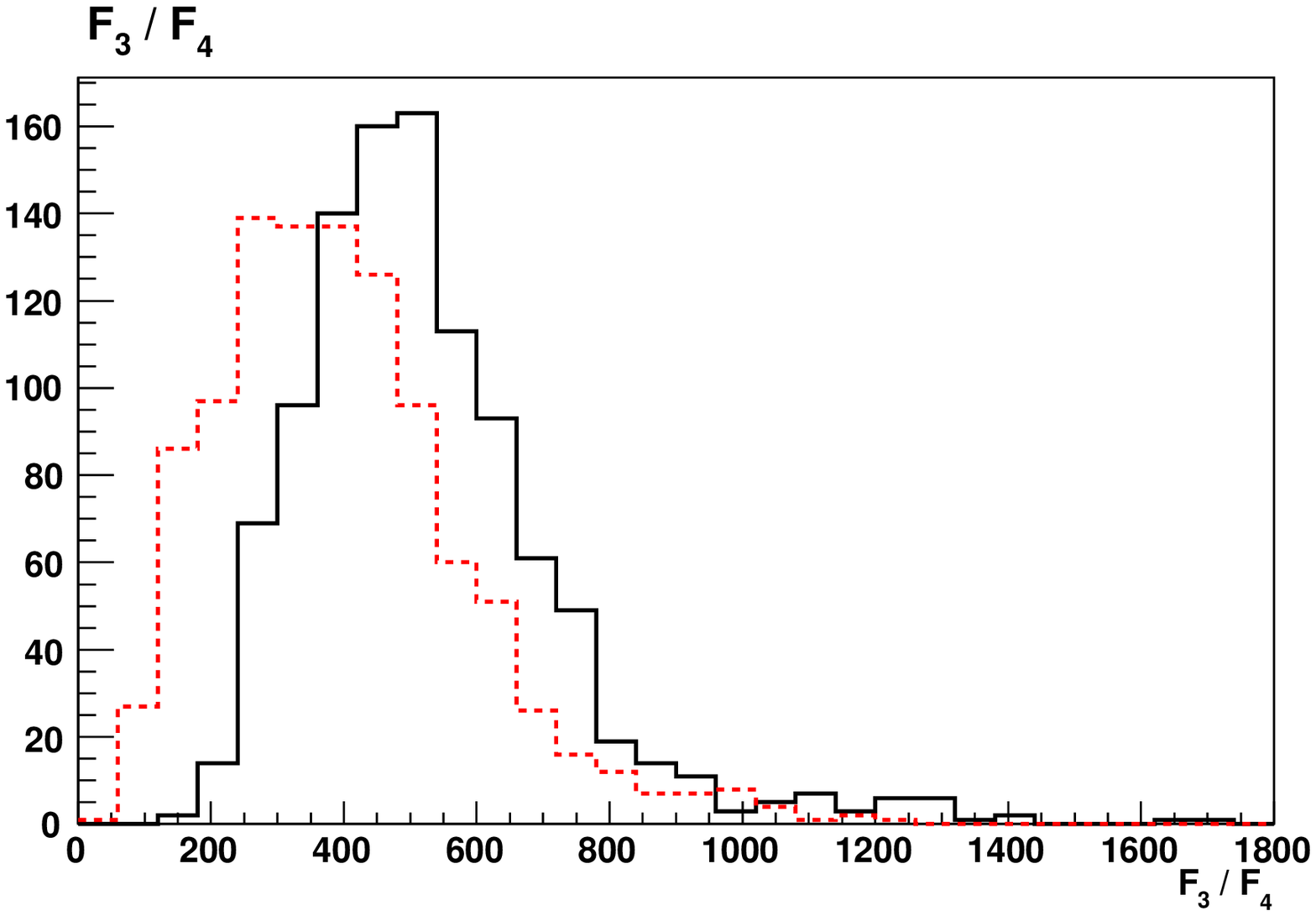}\\
\includegraphics[width=0.45\textwidth]{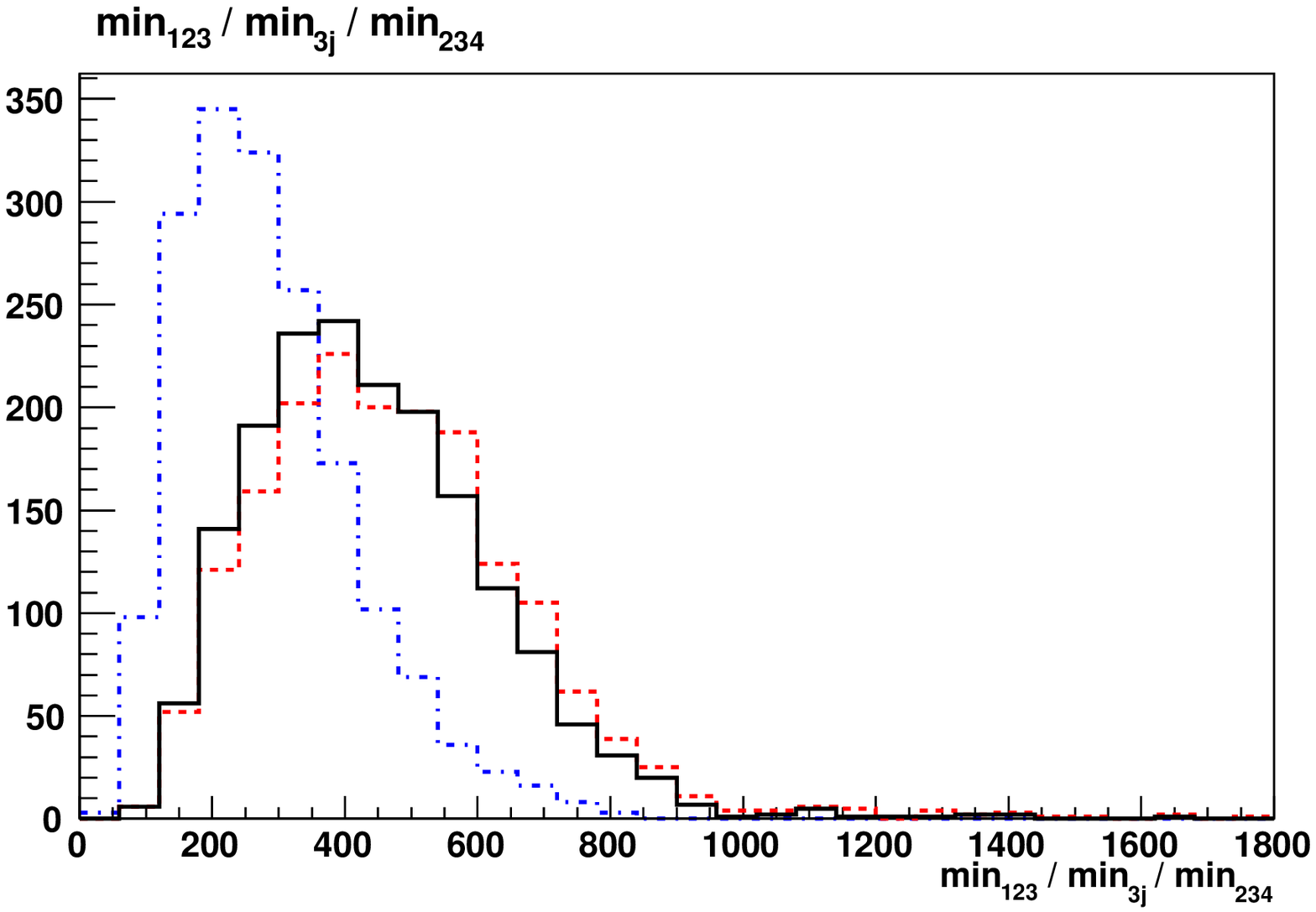}
\includegraphics[width=0.45\textwidth]{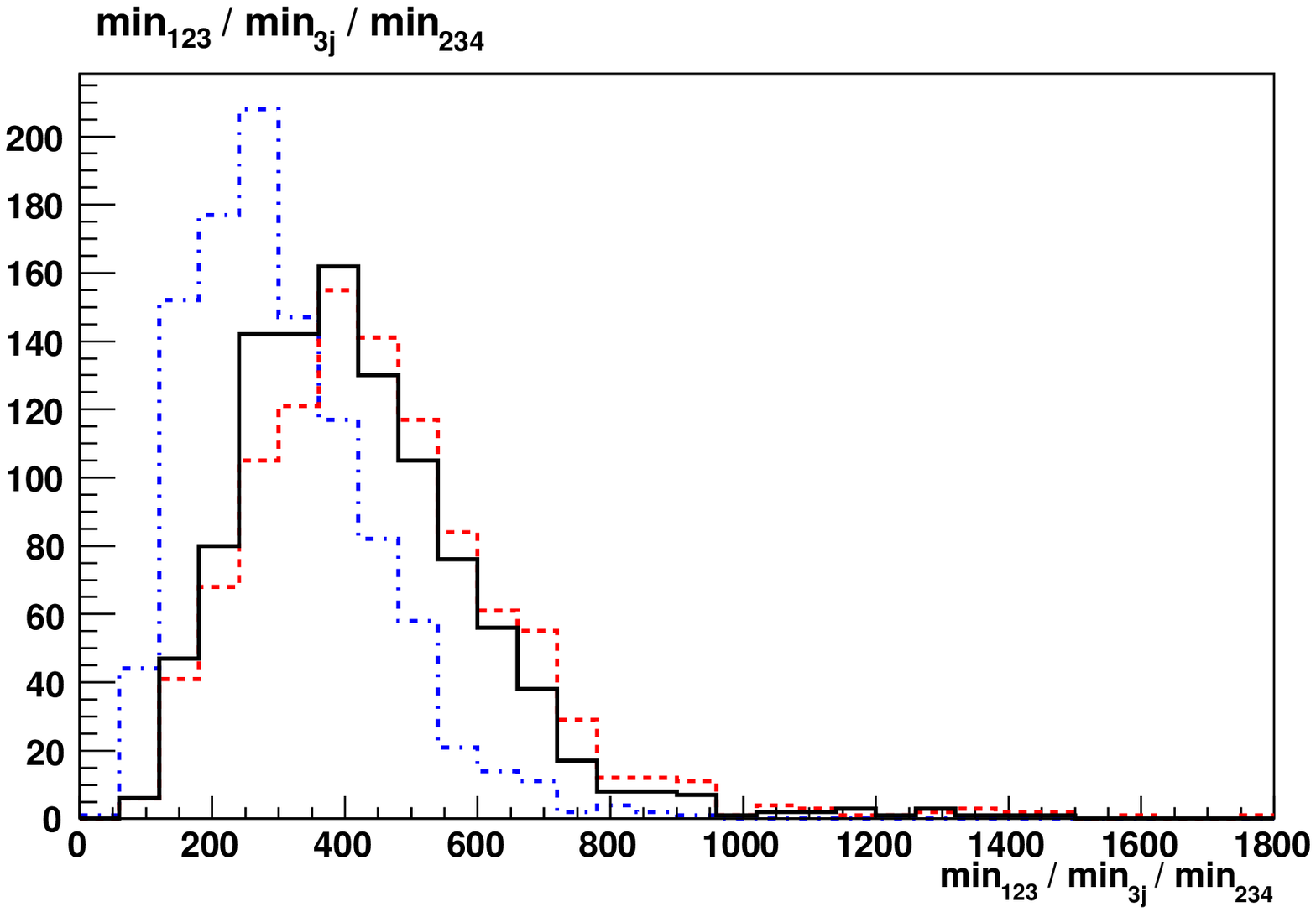}
\caption{Scenario C: performance of the variables under investigation. For details, see description of Fig.\ref{fig:mjet_typeA}.}
\label{fig:mjet_typeC}
\end{figure}

In the type C scenario, decoupled squarks with masses of 10 TeV are
not accessible at the LHC energy and the $\tilde g \tilde g$
production is the unique SUSY QCD production.  Many studies regarding
combinatorial issues are based on this type of spectrum, and most of
them consider only one particular gluino decay mode: $\tilde g \to jj
\tilde B$.

In Fig.\,\ref{fig:mjet_typeC}, we show the results for this
spectrum. As can be seen, almost all variables successfully recover the wino
edge at 800\,GeV. For the bino event selection, most variables suffer
from poor statistics in the vicinity of the true endpoint,
1000\,GeV.  It seems $F_4$, $min_{123}$ and $min_{3j}$ perform quite well
since their distribution linearly fall down to the true endpoint.


\section{Endpoint determination}

\begin{figure}[!t]
\centering
\includegraphics[width=0.45\textwidth]{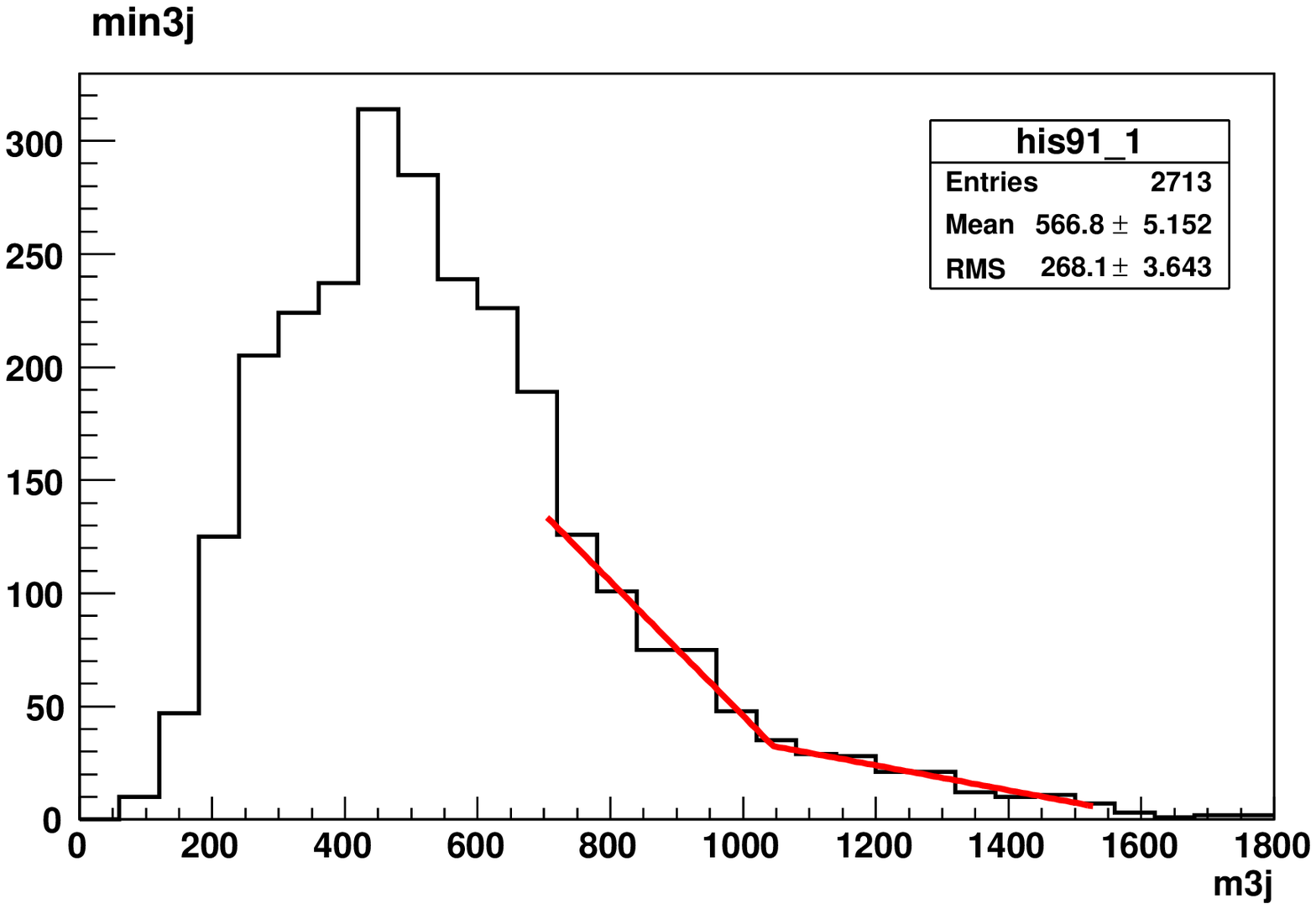}
\includegraphics[width=0.45\textwidth]{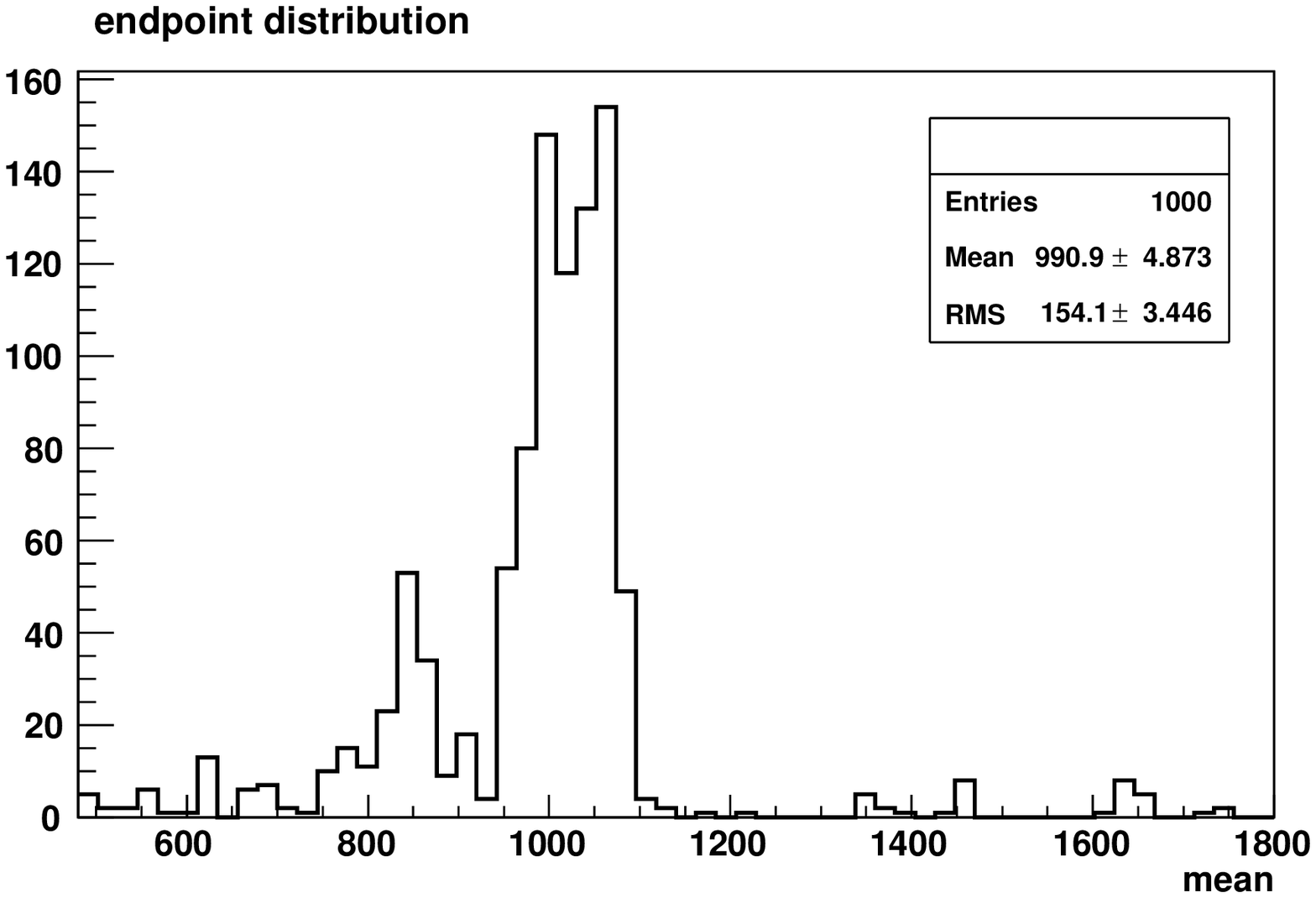}\\
\includegraphics[width=0.45\textwidth]{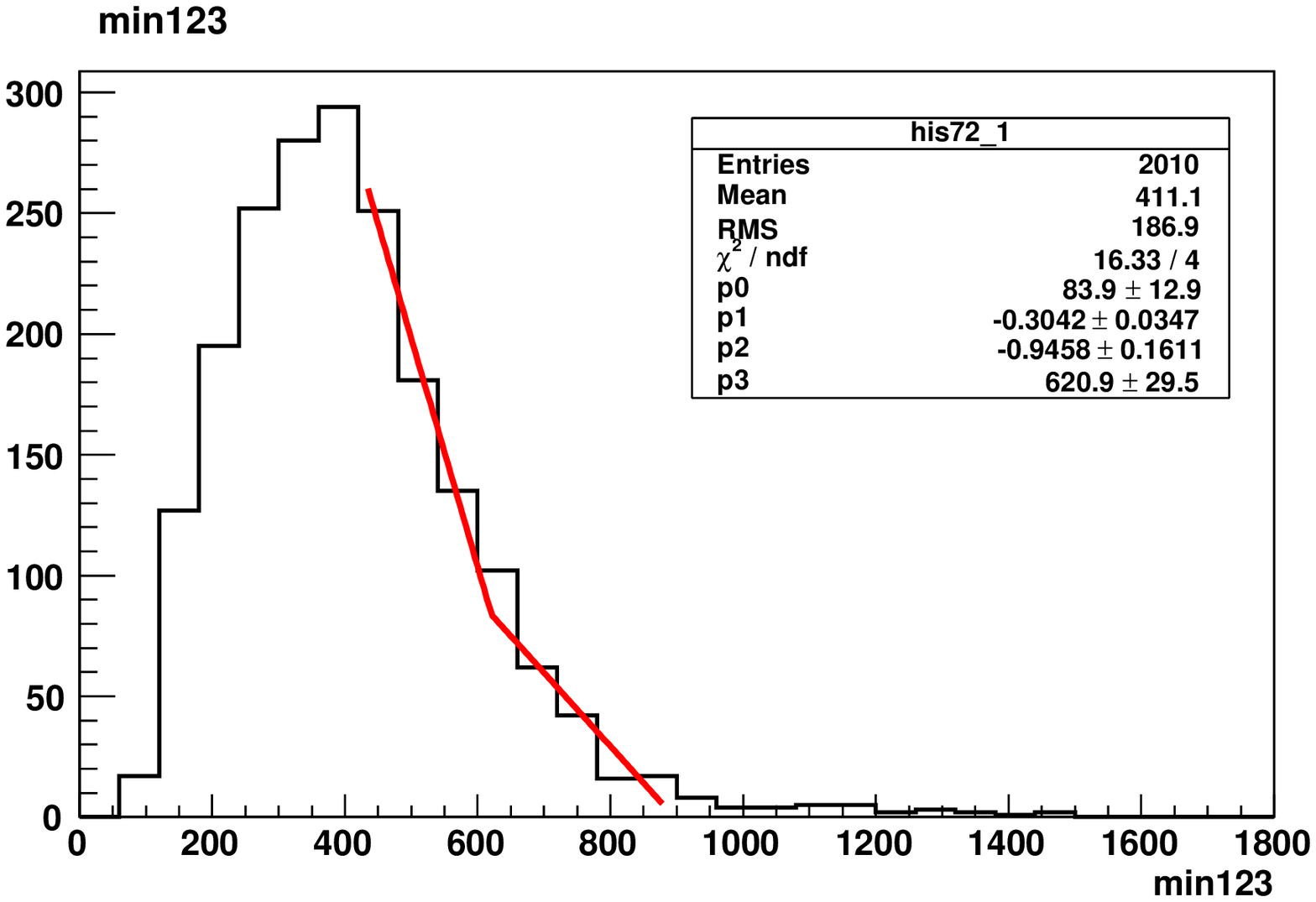}
\includegraphics[width=0.45\textwidth]{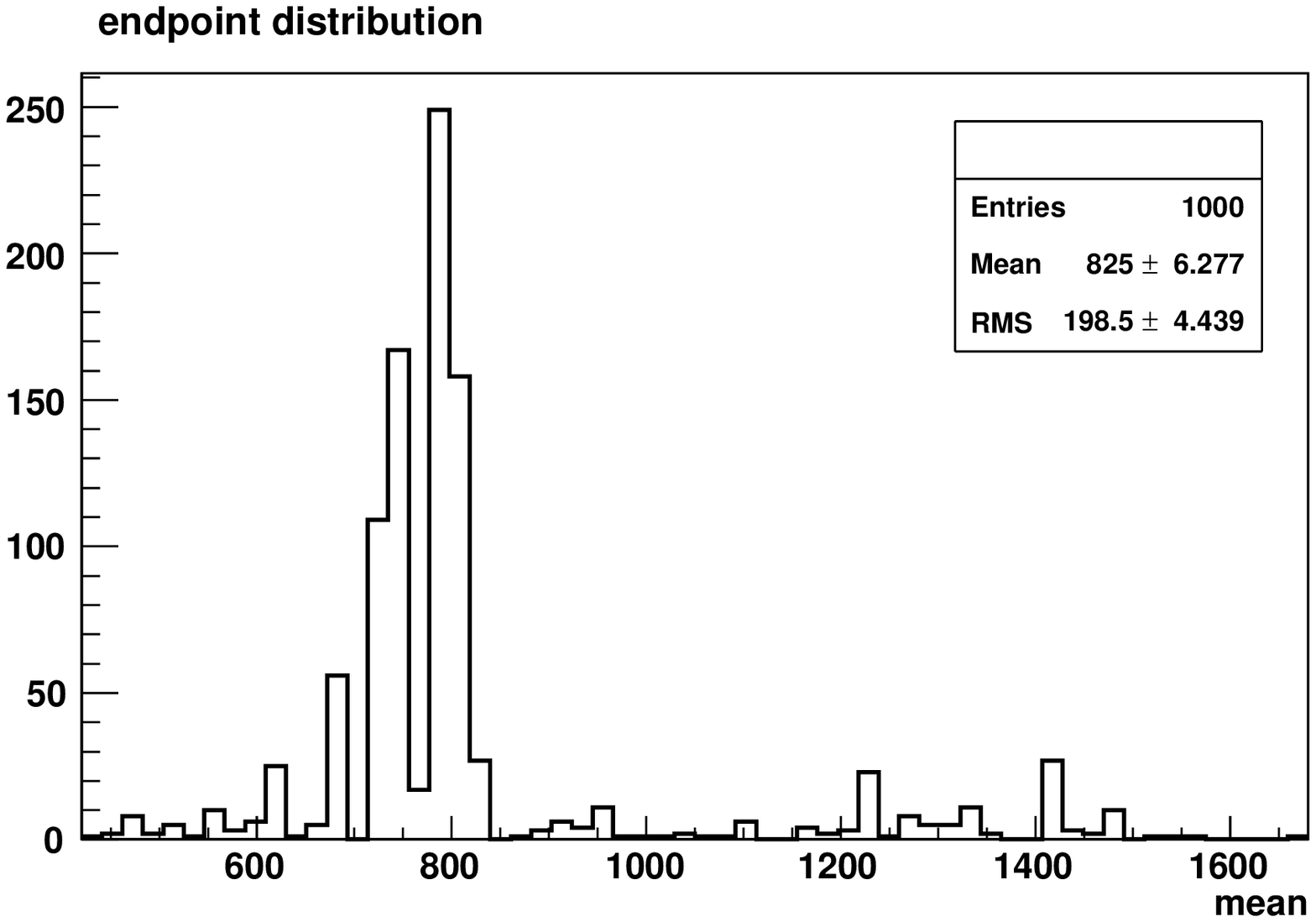}
\caption{Examples of bino and wino selection in scenario B and C, respectively: we use the edge-to-bump method to obtain a naive endpoint distribution of the variables $min_{3j}$ (bino, first row) and $min_{123}$ (wino, second row).}
\label{fig:mjet_fitexample}
\end{figure}

We attempt to estimate the endpoints of the distributions
quantitatively by adopting the edge-to-bump  
method proposed in \cite{Curtin:2011ng}. In that method, ``kinky''
features of distributions which could be originating from an
underlying kinematic feature of the distribution or simply be
artifacts of statistical fluctuations as well as trigger and cut
effects distorting originally smoother distributions, are turned into
bumps. Bumps are far easier accessible to data analysis methods, and it is easier to 
define a systematic error on the procedure of edge determination as this is 
translated to a statistical error of pseudo-experiments with different
edge-to-bump conversion parameters. Roughly speaking, the edge-to-bump 
method fits the changes in the slopes of the distributions and
translates them into peaks. The more distinct a peak is, the more
likely it is that there is a true kinematic kink in the original
distribution. In this method, for that purpose a naive linear kink fit
is used. 
The resulting value of the fit usually suffers from a
non-negligible correlation to the fitting range set by hand. To
minimize this artifact, we carry out 1000 naive kink fits with
randomly generated fitting ranges and obtain a statistical distribution of endpoint positions.

Fig. \ref{fig:mjet_fitexample} shows two examples of endpoint distributions.  
They are obtained by fitting two different distributions for the bino and wino selection in scenario B and C.
As we can see, the peak of the distribution is strongly correlated with the theoretically expected endpoint. 

As a first estimate one might give the mean value and the corresponding standard deviation as associated error. However, smaller peaks and non-negligible contributions far off the main peak lead to shifted mean values and considerably large errors ($\mathcal O(300)$ GeV). Thus to quantitatively estimate the actual endpoint position and get rid of the contributions far away from the main peak, we apply the following procedure:
\begin{table}[!t]
  \centering
\small
  \begin{tabular}{|c|c|c|c|c|c|c|c|}
    \hline
    endpt. & $min_{123}$ & $min_{234}$ & $min_{3j}$ & $m_{12}^{(1)}$ & $m_{12}^{(2)}$ & $F_3$ & $F4$ \\
    \hline
    \hline
    \multicolumn{8}{|c|}{scenario A} \\
    \hline
    bino & 1106 $\pm$ 52 & 570 $\pm$ 14  & 1125 $\pm$ 106 & 822 $\pm$ 21 & \textbf{ 1012 $\pm$ 104 } & 686 $\pm$ 33 & 1191 $\pm$ 132 \\
    wino & 908 $\pm$ 83 & 665 $\pm$ 34 & 948 $\pm$ 99 & 932 $\pm$ 31 & \textbf{ 780 $\pm$ 26 } & \textbf{ 794 $\pm$ 33 } & 1031 $\pm$ 53\\
    \hline
    \hline
    \multicolumn{8}{|c|}{scenario B} \\
    \hline
    bino & {\bf 986 $\pm$ 36} & 773 $\pm$ 147 & \textbf{ 1028 $\pm$ 34} & \textbf{1010 $\pm$ 6 }& 794 $\pm$ 49 & 766 $\pm$ 25 & 1046 $\pm$ 66 \\
    wino & 895 $\pm$ 23 &  \textbf{748 $\pm$ 68}  & 892 $\pm$ 18  & 958 $\pm$ 10 & \textbf{819 $\pm$ 47} & 911 $\pm$ 51 & 928 $\pm$ 37 \\
    \hline
    \hline
    \multicolumn{8}{|c|}{scenario C} \\
    \hline
    bino & 812 $\pm$ 24 & 545 $\pm$ 8 & \textbf{921 $\pm$ 37} & 816 $\pm$ 29 & 721 $\pm$ 90 & 708 $\pm$ 22 & \textbf{894 $\pm$ 57}  \\
    wino & 778 $\pm$ 23 & 577 $\pm$ 19 & \textbf{804 $\pm$ 6} & 769 $\pm$ 47 & 764 $\pm$ 14 & 708 $\pm$ 38 & \textbf{793 $\pm$ 7} \\
    \hline
  \end{tabular}
  \caption{Fit values for the discussed variables in the two endpoint selections, obtained with our own implementation of the edge-to-bump method \cite{Curtin:2011ng}. The values closest to the true endpoints are highlighted in bold face.} 
  \label{tab:fit_results}
\end{table}
\begin{enumerate}
\item calulate the mean value $\hat m_0$ and standard deviation $\sigma_0$ of the complete distribution
\item redefine the range of the distribution according to the above values: $\hat m_0 \pm 2 \sigma_0$
\item calculate a new mean value $\hat m_i$ and standard deviation $\sigma_i$ inside the range defined above
\item use $\hat m_i$ and $\sigma_i$ as a new \textit{seed} and start over with point 2
\end{enumerate}
Iterating three to six times the steps 2-4, we find convergence of $\hat m_i$ and $\sigma_i$. 
The resulting mean values and errors obtained by this procedure are listed in Table\,\ref{tab:fit_results}.
A first conclusion is that no variable works perfectly for all scenarios, which suggests that one should carefully choose the variables depending on the mass spectrum and 
the endpoints (bino or wino) to be measured.
In scenario A, $m_{12}^{(2)}$ serves the best estimates of the bino and wino endpoints among all variables.
$min_{123}$ and $min_{3j}$ possess shifts of about 100\,GeV towards higher mass regions but preserve 
the correct difference between the wino and bino edges. 
In scenario B, the $\tilde q \tilde g$ associated production gives an additional high $p_T$ jet and the size of the combinatorial background
is the largest among the three scenarios.  
However, $min_{123}$, $min_{3j}$ and $m_{12}^{(1)}$ ($min_{234}$ and $m_{12}^{(2)}$) provide consistent results with the theoretically
expected value for the bino (wino) edge measurement and the errors are somewhat smaller than in scenario A. 
On the other hand, the difference among the two endpoints are underestimated by most variables. 
This reflects the importance to use an appropriate variable depending on wether a wino or bino selection criterium is applied.
In scenario C, all variables tend to underestimate the bino endpoint, despite the fact that
the combinatorial and SUSY backgrounds are smallest in size among the three scenarios.
This lower bias effect stems from poor statistics due to a small SUSY cross section in scenario C.
We checked that the tendency of underestimation is removed when the number of events is increased artificially. 
For the wino edge, many variables, $min_{123}$, $min_{3j}$, $m_{12}^{(1)}$, $F_4$, give good results with small errors. 

Finally we want to stress that the quoted errors in Table\,\ref{tab:fit_results} are only errors originating from the fitting range dependence on fit results and there exist other sources of errors, which should be taken into account.
For example, the statistical error on each bin content and the biases inherent in the according variable as well as backgrounds would all give contributions of the size of the errors we quoted.
A careful estimation of these is beyond the scope of this paper but will be an important subject for gluino endpoint measurements.

\section{Conclusions}
In this work we studied the feasibility of the
gluino dijet mass edge measurement in a realistic LHC
environment. This includes both full SUSY backgrounds and
combinatorical miscombinations of particles, as well as 
effects of initial and final state radiation and finite detector
resolution. Several methods in the literature explicitly rely on the
precise knowledge of a particular endpoint to be able to access
information on masses in decay cascades, to resolve combinatorial
issues, or determine another kinematical variable. Often QCD radiation
and detector effects have been neglected in first phenomenological
investigations. By utilizing considerations from the analysis of
topological configurations of gluino decay cascades at the parton
level, we find that the surviving correlation between the number of
parton level and detector level jets is sufficient in distinguishing bino and wino endpoints of gluino decays with semi-inclusive jet multiplicities and the use
of leptons as further selection criteria. To assess the impact of
different mass spectra, we analyse three distinct (semi-)simplified
models: two with small and large mass differences between gluino and
squarks, and one scenario with decoupled scalars. In these models, we
make use of existing kinematic variables and propose new ones, where necessary,
that reduce the combinatorical problem and, when applied to
preselected events, allow for the excavation of two distinct gluino
endpoints. In 
general, the kinematic variables presented here are robust
against contaminations from QCD radiation, underlying (non-signal)
SUSY processes as well as detector effects. The resulting
distributions are fitted with the so-called edge-to-bump method
minimizing the artifact of the fit. These first estimates of the
gluino dijet endpoints are mostly consistent with their expected values and
thus proof the validity of our method. Hence, it seems 
possible to measure the gluino dijet edges for basically all cases
where the squarks are heavier than the gluino, a parameter space that
seems to be favored by recent Higgs search analyses from the LHC
experiments.


\subsubsection*{Acknowledgements}

We would like to thank Christian Sander and Matthias Schr\"oder for
valuable comments and helpful discussions. JRR and DW like to thank the Institute of Physics
of the University of Freiburg for their hospitality. 


\end{document}